\documentclass{aastex}
\usepackage{emulateapj5, psfig, epsfig}

\slugcomment{accepted by The Astronomical Journal, July 31, 2001}

\shorttitle{Turnoff Ages of NGC~288 and NGC~362}
\shortauthors{Bellazzini et al.}

\begin{document}

\title{Age as the Second Parameter in NGC~288/NGC~362? \\
       I. Turnoff Ages: a Purely Differential Comparison}

\author{Michele Bellazzini, Flavio Fusi Pecci\altaffilmark{1}
Francesco R. Ferraro, Silvia Galleti}
\affil{Osservatorio Astronomico di Bologna, Via Ranzani 1, 40127, Bologna, 
ITALY}
\email{bellazzini@bo.astro.it, flavio@bo.astro.it, 
ferraro@bo.astro.it, l\_galleti@bo.astro.it}

\author{M\'arcio Catelan\altaffilmark{2}}
\affil{University of Virginia, Department of Astronomy, 
        P.O. Box 3818, Charlottesville, VA, 22903-0818 }
\email{catelan@virginia.edu}

\and

\author{Wayne B. Landsman}
\affil{Raytheon ITSS, NASA/GSFC, Greenbelt, MD 20771, USA}
\email{landsman@mpb.gsfc.nasa.gov}

\altaffiltext{1}{Stazione Astronomica di Cagliari, Loc. Poggio
dei Pini, Strada 54, 09012 Capoterra (CA), ITALY}
\altaffiltext{2}{Hubble Fellow}


\begin{abstract}
We present deep $V$,~$I$ photometry of the globular clusters NGC~288, NGC~362 and 
NGC~1851 obtained during a single observational run under strictly homogeneous
conditions. We use the bimodal horizontal branch (HB) of NGC~1851 
as a ``bridge'' to obtain the optimum relative match between the HBs of 
NGC~288 and NGC~362. 
In this way we can effectively remove the uncertainties associated with distance,
reddening and inhomogeneities in the absolute calibration, thus obtaining a
very robust, purely differential estimate of the age difference between these
two clusters. According to the bridge test, NGC~288 is found to be older than 
NGC~362 by $2\pm 1$ Gyr. This result is fully confirmed also by all classical 
differential age diagnostics, either based on the luminosity 
($\Delta V_{\rm TO}^{\rm HB}$) or color [$\delta (V-I)_{@2.5}$] of the 
main sequence turnoff point.
 
\end{abstract}



\section{Introduction}

The ``Second Parameter Effect'' \cite[SPE; see][and references therein]{fb97}  
has been at the center of the debate on Galaxy formation during the last two
decades. In particular the interpretation of {\em most} of the observed 
differences of HB morphology at fixed metal content in globular clusters
(GCs) in terms of age differences \citep{sz78,ldz94,lee99} leads 
to the conclusion that a large age spread is present in the stellar 
Galactic halo. 
In the last few years a body of stringent observational evidence has
demonstrated that this interpretation was unjustified
\citep{svb96,har97,buo98a,buo98b,ros99,cat2000,cfr00,vdb00,maxt01}. 
Moreover the advent of
larger and/or space located telescopes and more efficient detectors has removed
any need \cite[e.g.,][]{sz78}
of relying on HB morphologies to estimate ages, carrying within reach
the best suited age indicator 
\cite[i.e., the main sequence turnoff (MSTO), see][]{rf88,svb96} over the 
whole range of distances of Galactic GCs.

We are now in the much safer condition of
splitting the old version of the SPE conundrum into two distinct and well
defined questions: (1) what physical quantities are actually driving differences
in HB morphology, i.e. a problem of stellar astrophysics, and (2) what are the
age distribution, the age-metallicity and the age-galactocentric distance
relations of the Galactic GC system, i.e. a problem of Galactic astronomy.

In the context of the SPE debate, the classical GC pair 
NGC~288 and NGC~362 has been the subject of extensive study and long standing 
controversies \cite[see][hereinafter SVB96, and references therein]{svb96}. 
In this paper we present the results of a specific observational test devised 
to measure the age difference between these two clusters with high accuracy.
In doing this, we take also advantage of the detailed abundance analysis 
recently published by \citet{sk00}.

The results of the test will drive a detailed investigation on the origin of the
observed difference in HB morphology, performed with modern 
color-magnitude diagram (CMD) synthesis techniques  
\cite[as adopted by][]{cat2000,cfr00}, that  will be presented in a companion 
paper \cite[][hereafter Paper~II]{pap2}.

The plan of the paper is as follows: in \S2 we briefly recall and comment on
previous results on the age differences between NGC~362 and NGC~288, and we
describe the planned test.
The metal abundance estimates found in the literature are also reviewed.
In \S3 the observational material and data reductions are
described and \S4 is devoted to the actual application of the age difference
test. A direct comparison with previous realizations of the bridge test is also
presented and it is demonstrated that the present application is superior to 
previous ones (\S4.5).  
The underlying assumptions at the basis of the bridge test are also critically 
discussed (\S4.6). In \S5 we comment on an interesting spin-off of our
results concerning the origins of NGC~288 and NGC~362.
Finally, the results of the whole analysis are summarized and 
discussed in \S6.

\section{NGC~288/NGC~362, a Controversial Couple}

NGC~288 and NGC~362 are two well known and relatively nearby southern clusters
\cite[$(M-m)_0=14.73$ and $(M-m)_0=14.68$ respectively, ][]{fer99}.
Since the early spectrophotometric estimates 
\cite[][and references therein]{zinn80} it was realized that they were quite 
similar in metal content, a claim that has been confirmed many times 
(see \S2.1.). 
Hence, the strong difference in HB morphology [$(B-R)/(B+V+R)=0.95\pm 0.08$ 
for NGC~288 and $(B-R)/(B+V+R)=-0.87\pm 0.08$ for NGC~362 
\citep{ldz94}\footnote{Where $B$, $R$ and $V$
are the number of stars bluer than the instability strip ($B$), redder than the
instability strip ($R$) and lying inside the instability strip, i.e. RR Lyrae
variables ($V$).} at similar metallicity made these two clusters an excellent
test case to study the SPE.

Based on the HB morphology,
a large difference in age ($> 7$~Gyr) was suggested as the SP at work in this 
couple by \citet{ldz88}, later revised by the same authors to $\sim 3-4$ Gyr
\citep{ldz94}. \citet{cat93} found that age differences larger than  
$\sim 3$~Gyr were indeed required to account for this second parameter pair, 
unless absolute ages smaller than $\sim 10$~Gyr were assumed. 
Hence it is important to obtain independent age estimates based 
on observations of the MSTO, to settle the issue whether 
or not age can be considered the (sole) ``second parameter" at play in this 
case (see Paper~II for an extensive discussion).

While the comparison of two clusters of similar metal content remains the case
in which the most secure estimates of age differences can be obtained, a number
of important sources of uncertainty can still plague the measures 
\cite[][SVB96]{bol89}:

\begin{enumerate}

\item If the considered clusters have similar HB morphologies, the
match of the HBs obtained by shifting of the CMDs can provide a direct
comparison virtually independent of distance and reddening. In the opposite case,
i.e. very different morphologies, as for NGC~288 and NGC~362, this approach is
obviously impossible and independent estimates of the relative distance and
reddening are necessary. These estimates are the major contributors to the final
error budget of the relative age measure \citep{rf88,bol89}. In particular, the
estimate of the relative distance has to be based on (a) questionable
extrapolations of the observed HB to an {\em unobserved} common level 
(the theoretical zero-age horizontal branch (ZAHB) at the instability strip 
level, for example) or on (b) matching of the main sequences, 
made difficult by the morphology of the sequence (i.e. its high slope in 
the CMD) and by uncertainties in the reddening \citep{bol89};

\item Even if the overall metal abundance is similar, undetected (or unmeasured)
differences in $\alpha$-elements abundances or in 
primordial He abundance ($Y$)
can lead to significant misinterpretation of observed differences in
the location of the MSTO and the HB \cite[][SVB96]{rf88}. 
A detailed and comparative 
abundance analysis of a significant sample of NGC~288 and NGC~362 stars has become
available only very recently \citep{sk00}. The problem of He abundance will be
briefly discussed in \S2.1;

\item The comparison of photometric material taken under  different conditions
(i.e. telescope, cameras, observing run, set of absolute calibrators, actually
adopted filters, etc.) is not necessarily safe and can introduce significant
errors in age difference estimates \citep{bol89}. 
For instance, slight differences in the
adopted filters can introduce color equations depending on the local set-up of
the observations. Merging deep and bright photometries taken from
different sources can introduce deformations in the CMD, altering the final
comparison between the MSTO, for example by changing the relative difference
between the MSTO and the HB magnitudes and colors in the CMD. 

\end{enumerate}

While many authors have provided estimates of the age difference between NGC~288
and NGC~362 based on sufficiently deep photometry to sensibly measure the MSTO,
their analyses suffered from at least one of the problems described above.

\citet{pjh87} compared their deep CCD photometry of NGC~288 with a composite 
CMD of NGC~362 obtained by merging the faint stars sample observed with a CCD 
camera
by \citet{bol87} with the bright stars sample obtained by \citet{har82} based on
photographic plates calibrated with photoelectric photometry. Hence, this
analysis was prone to all three sources of uncertainties described above. 
The adopted difference in apparent distance modulus was 
$\Delta \mu = (m-M)_V^{NGC 362} - (m-M)_V^{NGC 288} = +0.10$.
The Pound et al. results were compatible with a significant age difference 
between the two clusters, NGC~288 being older.   
However they concluded that the age
difference was not sufficient to explain the observed HB morphologies and 
suggested the existence of as yet undetected differences in chemical 
composition.  
 
Both \citet{bol89} and \citet{GN90} tailored their observations to minimize the
effects associated with point 3, by observing both clusters in the same
observing run, with the same observational setup and tying the photometry
to a common calibration \cite[see the discussion in \S6 in][]{bol89}. 
However, they were forced to assume and/or derive
the relative distance and reddening from less safe ways than direct HB matching,
thus both studies suffer from the associated uncertainties (point 1).  
The two analyses reached the same conclusion, i.e. that NGC~288 is $\sim 3$~Gyr 
older than NGC~362. The adopted $\Delta \mu$ were $+0.10$ for \citet{bol89} and
$+0.07$ for \citet{GN90}.
 
Age indicators based on color differences between the MSTO and some point at 
the base of the red giant branch (RGB) 
[$\Delta(\bv)_{\rm TO,RGB}$] are by definition independent 
of distance and reddening. In
case of application to clusters of the same metallicity, their strong 
dependence
on this last parameter can be ignored 
\cite[see the discussion in SVB96 and][]{buo98a}. 
Furthermore, given the vicinity in the CMD of the two involved
features, strong influences by inhomogeneous calibrations are less likely.
Their most noticeable drawback stands in the theoretical calibration, since, as
well known since long ago \citep{rf88,buo98a}, uncertainties in the free 
parameters of the stellar models (as, e.g., the mixing length) and in the
transformations from the theoretical ($L,T_{\rm eff}$) to the observational 
($V$, \bv) plane affect model predictions about {\em colors} much more than 
those about luminosity. This is particularly critical because 
of the exceptional sensitivity of these kinds of {\em horizontal}\footnote{We
refer to differential  age indicators as {\em horizontal} if they are
defined as color differences, and {\em vertical} if they are defined as
magnitude differences; see SVB96.} parameters to age 
\citep{buo98a}. For instance, in the
calibration adopted by \citet{sd90} to analyze the NGC~288 - NGC~362 pair by
means of a horizontal differential parameter, an age difference of 1 Gyr is
associated to a difference in the adopted observable of 0.01 - 0.02 mag, 
depending on
the absolute age of the oldest cluster. \citet{sd90}, based on unpublished data
from other authors, concluded that NGC~288 is $3.1 \pm 0.9$ Gyr older than NGC
362, but the formal error on their 
$\Delta(\bv)_{\rm NGC~288} - \Delta(\bv)_{\rm NGC~362}=0.05$
is 0.014 mag, i.e. $\sim 30 \%$ and the uncertainty associated with the
theoretical calibration is not quantified. 

Independently of the quoted problems affecting each of the described estimates,
similar results were found by the different authors and the question appeared as
almost settled in the middle of the '90s, when SVB96 introduced a 
decisive change of perspective. 
These authors were the first to apply the idea of using a cluster with a bimodal 
HB morphology and similar metallicity (NGC~1851) to perform a purely 
differential match between NGC~362 and NGC~288, thus removing the uncertainties
associated with point 1, while performing an age test based on the whole
morphology of the MSTO region, including magnitudes and colors. 
Hereafter we will refer to the test devised by SVB96 as the
{\em bridge test}, since the HB of NGC~1851 is used as a bridge to match the
HBs of NGC~288 and NGC~362 \citep{vdb00}.
In particular 
SVB96 matched the MSTO regions of NGC~362 and NGC~288 to that of NGC~1851
and then checked whether the HB of NGC~362 matched the red part of the HB of 
NGC~1851 and if the HB of NGC~288 matched the blue HB of NGC~1851. 
Finding a good overall
match they concluded that `` ... all three clusters do, indeed, have the same
age to within quite a small uncertainty ($\le 1$ Gyr). Moreover, small
cluster-to-cluster differences in [Fe/H] or $[\alpha/{\rm Fe}]$ will not alter this
conclusion because we have effectively used the $\Delta V$ (vertical parameter) 
method, which is insensitive to modest changes in heavy element abundances
...''. 
  
While the comparison devised by SVB96 appears to be the most robust 
either from an observational or a theoretical point of view, their actual
application of the test suffers from all the uncertainties described in point 3
above. In fact, they adopted the photometry of \citet{w92} for NGC~1851, the
data of \citet{berg93} for the bright ($V<17.5$~mag) part of the NGC~288 CMD and
\citet{bol92} for the faint part, and finally the photographic photometry of 
\citet{har82} for the
bright ($V<17$~mag) 
part of the NGC~362 CMD, and \citet{vdb90} for the faint part 
of the same cluster. Very inhomogeneous databases were compared, therefore
both internal (bright vs. faint) and external (differences in the absolute
calibrations and/or observational setup) effects can undermine the test.
Furthermore the plots of the CMD of NGC~1851 erroneously included a few RR Lyrae
among the brightest blue HB stars, an occurrence
that makes less stringent the matching between the HB of NGC~1851 and the 
one of NGC~288 (see the {\em note added in proof} in SVB96). This source of
uncertainty was removed in the analysis by \citet{vdb00} who confirmed 
the SVB96 result (the derived $\Delta \mu$ are $-0.13$ for SVB96 and $-0.15$
for \citet{vdb00}). Nevertheless, VandenBerg adopts the same inhomogeneous 
dataset as SVB96. The pernicious effects of the use of heterogenous databases 
in the previous realizations of the bridge test will be described and 
discussed in \S4.5.
It is straightforward to conclude that the SVB96 test can be refined by
settling the point-3-related problems and by adopting a slightly different
strategy. In our view, a more
tightly constraining test would be obtained by {\em matching the HBs} and 
comparing the resulting differences in the MSTO regions instead of the
opposite. For instance, the clumpy red HBs of NGC~362 and NGC~1851 provide a 
much stronger reference point for the comparison than the MSTO-SGB 
almost-sinusoidal curve.

There are two recent studies in the literature that attempt a relative age 
estimate involving the three clusters we are dealing with and that present a 
high degree  of homogeneity in the observational material. 
\citet{ros99,ros00a} presented a
homogeneous and accurately calibrated photometric database of southern GCs 
including NGC~288, NGC~362 and NGC~1851. They find an age difference 
between NGC~288 and NGC~362 of $2.6 \pm 1$ Gyr and of $2.2 \pm 1$ Gyr between
NGC~288 and NGC~1851, NGC~288 being the older cluster. 
The final CMDs are more fuzzy and contain less
stars than our dataset [see in particular the scarcely populated 
HB of NGC~1851, Fig.~8 of \citet{ros00a}]. In the analysis, \citet{ros99} did 
not attempt a specific bridge test, since they were interested 
in the establishment 
of a global relative age scale based on both vertical and horizontal age 
diagnostics. 
\citet{gru99} presented preliminary results of a bridge test involving the
above quoted clusters based on Str\"omgren $uvby$ photometry, carried out with 
the same instrumental set-up.
He applied the
test in the same way as SVB96 and \citet{vdb00} and confirmed their results.
A deeper analysis (still based on the assumption that the three clusters
have the same abundance of heavy elements)
however showed that an age difference up to $ 2$ Gyr between 
NGC~288 and NGC~362 cannot be entirely excluded 
(Grundahl, private communication).

From the above discussion it is clear that (a) despite extensive efforts the
question of the age difference between NGC~288 and NGC~362 is not 
settled yet, and (b) a bridge test performed with homogeneous observational 
material is - presently - the most secure way to obtain the final answer
(see \S4.5).
With this purpose we obtained {\em deep $V$ and $I$ photometry of NGC~288,
NGC~362 and NGC~1851 during the same observational run, with the same
observational setup and locked to the same calibrating color equations}. 
We used this observational material to perform a more robust version of the
bridge test, obtaining the best minimization of the uncertainties related to
points 1 and 3 above, given the current technical limitations.  The next
subsection will deal with point 2.

\subsection{Chemical composition}

Any of the above described estimates of the relative ages of NGC~288 and NGC~362
have to take into account the following question: do indeed the two clusters have
the same chemical composition? While photometric and spectroscopic estimates
were in agreement since the earliest times, there was room for non-negligible
differences in the overall metal content 
\cite[up to $\sim 0.5$~dex;][]{zinn80,zinn85}
and, above all, the $\alpha$-element abundance as well as the extent of 
possible mixing phenomena \citep{gra00,swc} were largely unconstrained. 
As a reference we
report the metallicity listed by \citet{zinn84}, i.e. 
$[{\rm Fe/H}]=-1.40\pm 0.12$ for NGC~288 and $-1.27\pm 0.07$ for NGC~362.

Studies based on high resolution spectra \citep{ps83,gra87,dick91,cr} 
resulted in detailed analysis of the abundance pattern 
which confirms a close similarity between the two clusters 
\cite[see][for references and discussion]{sk00}.

The very recent study by \citet{sk00} couples the accurate reconstruction of the
abundance pattern performed by \citet{gra87} (including also Al, Sc and Eu) with
the ``large'' sample approach by \citet{dick91}. They observed 13 and 12 HB
stars in NGC~288 and NGC~362, finding $[{\rm Fe/H}]=-1.39 \pm 0.01$ and 
$[{\rm Fe/H}]=-1.33 \pm 0.01$ respectively, and very similar abundance patterns 
(in particular, both clusters have $[\alpha/{\rm Fe}]\sim +0.3$).
They also analyzed possible diagnostics of mixing and mass loss and finally
concluded:{\em ``...NGC~288 and NGC~362 have roughly the same abundance, 
roughly the same $\alpha$ enhancement ratio, roughly the same percentage of
mixed stars, the same extent of deep mixing within those stars, and no extreme
mass loss taking place on the giant branch...''}\footnote{ We note here that, 
in spite of the unprecedented large survey of spectroscopic mass loss 
indicators in giant stars in NGC~288 and NGC~362
by \citet{sk00}, their sample is still inadequate to study differences in 
mass loss rates between giants in these two clusters. This is evident from Fig.~5
in their paper, where one clearly finds that the vast majority of their brighter 
studied stars are members of NGC~362, whereas their NGC~288 sample is much 
fainter. Therefore, the regime where more extreme mass loss might be expected 
for NGC~288 was simply not covered in the Shetrone \& Keane investigation, and 
the question whether NGC~288 giants may lose more or less mass than NGC~362 
giants close to the tip of the red giant branch remains open.}  

The only small difference pointed out by Shetrone \& Keane is in the oxygen 
abundance. From the analysis of the Na-O
anticorrelation (observed in both clusters) they argue that NGC~288 has a
primordial oxygen abundance larger by $\sim +0.15$ dex than NGC~362. Given the
current uncertainties \citep{mcw97} we regard this difference as marginal, 
in the present context.
However it has to be noted that, if real, it would provide a further problem in
the explanation of the difference in the HB morphology, since higher O
abundances have to correspond to redder morphology, all other parameter being
fixed (see Paper II for discussion). 

Despite some disagreement in the zero points of the metallicity
scale,\footnote{Inhomogeneity between different metallicity scales is a well
recognized problem affecting many fields of astrophysics  
\citep{jurcs,cg97,rutl97,mcw97}. See Paper~II for further discussion.} 
there is a broad consensus on the {\em similarity} 
in chemical composition between NGC~288 and NGC~362. It is difficult
to find another couple of clusters for which such a large and detailed amount of 
spectroscopic analysis is available showing that indeed they are a SPE couple,
i.e. they have the same metal content. Hence our age test is now grounded on the
most solid basis attainable with ``state of the art'' abundance analysis
techniques.

The case of NGC~1851 is much less fortunate since no high dispersion analysis is
available in the literature. All papers we have consulted adopt 
$[{\rm Fe/H}]=-1.33\pm 0.1$ from \citet{zinn85}. \citet{rh87} measured 
equivalent widths of Ca {\sc ii} lines in eight HB stars in NGC~1851, obtaining 
$[{\rm Fe/H}]=-1.4\pm 0.15$, in good agreement with the previous estimate. Thus we can
only conclude that NGC~1851 has an overall metallicity similar to NGC~288
and NGC~362. However this can be accepted as a sufficient condition for using
the CMD of this cluster as a bridge since it is difficult to conceive a
inhomogeneity in chemical composition that could change the relative luminosity
level of its blue and red HB in different ways 
 \cite[but see][and \S4.6 for different viewpoints and caveats]{swc}.

There is still an important factor in the chemical composition of the considered
clusters that may affect both age estimates and HB morphology and
cannot be fruitfully constrained with spectroscopy, i.e. the helium abundance 
$Y$.
\citet{san00} has recently reviewed the methods to determine $Y$ in GCs
and has concluded that the technique based on population ratios 
\cite[the so-called $R$ method,][]{iben} remains the most reliable one, 
despite the significant uncertainties associated.
The $R$ method has been applied to NGC~288, NGC~362 and NGC~1851 by several 
authors \citep{san00,zoc,buz,mary} adopting slightly different techniques. 
All the above quoted studies have not found any evidence of a difference 
in $Y$ between the three considered clusters, and the application of the $R$
method to our data confirms this result.
It has however to be recalled that the current uncertainties still 
leave room for the possibility of significant differences in the He abundance.

Finally, a recent analysis by \citet{bono} suggests that it is very unlikely
that early deep mixing phenomena may significantly affect galactic globulars.

\section{Observations and data reduction}

The observations have been carried out during the nights of 1997 January 
2 and 3, at the 2.2 m ESO/MPI telescope at La Silla
(Chile) with the EFOSC2 camera\footnote{See 
{\tt
http://www.ls.eso.org/lasilla/Telescopes/360cat/efosc/html/efosc2\_GENE.html}}, 
equipped with a Loral/Lesser $2048 \times 2048$ pixels CCD. 
The pixel scale is $0.26$
arcsec/pixel, and the effective field of view is $8 \times 8$ arcmin$^2$.
The gain is $1.63~{\rm e}^{-}/{\rm ADU}$ and the read-out noise is 
$6.3~{\rm e}^{-}$ rms. 

The seeing conditions were average during the January 2 night ($\sim 1.3$~arcsec)
when the NGC~1851 and NGC~288 observations were carried out, and worsened 
($\sim 2$~arcsec) during the second night when we observed NGC~362. 

For each cluster two partially overlapping fields have been observed:

\begin{itemize}

\item An inner field (INT), centered on the center of the cluster and observed 
with short and intermediate exposure times (1~s, 1~min, 2~min)
with the aim of sampling the bright stars.

\item An outer field (OUT) in which also long exposures (10~min) have 
been acquired in order to sample the main sequence in relatively uncrowded 
regions.

\end{itemize}

The rationale of the observational strategy was to obtain a large and complete
sample of evolved stars to have a well defined HB, and to obtain deep 
photometry in the most favorable conditions to get a clean MSTO region and a 
well sampled MS.

Each frame has been corrected for bias and flatfield and the overscan area has 
been trimmed using standard IRAF packages. 

\begin{figure*}[ht]
\figurenum{1}
\centerline{\psfig{figure=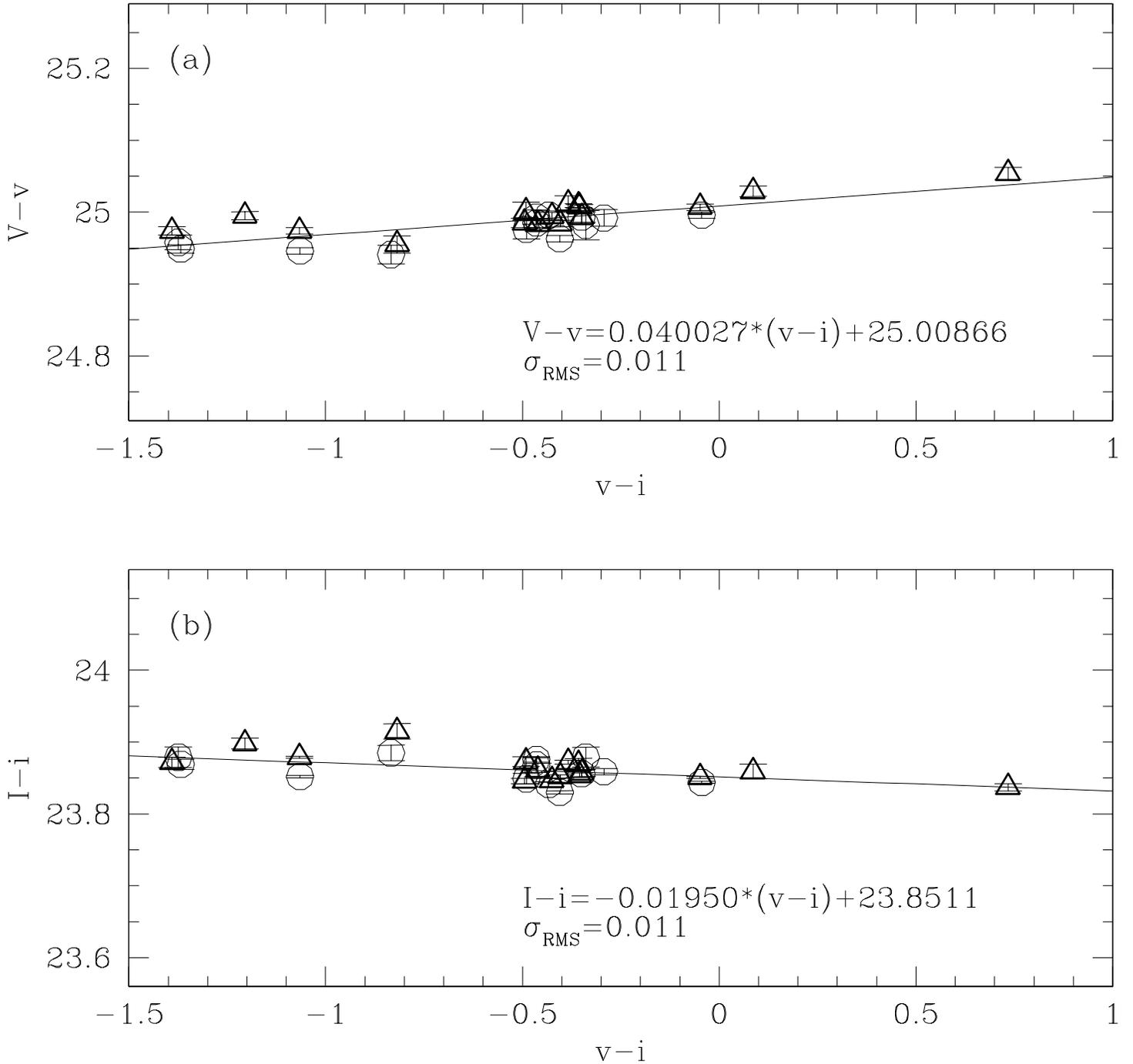}}
\caption{Differences between tabulated magnitudes ($V$, $I$) and instrumental 
magnitudes ($v$, $i$) versus instrumental color index ($v-i$) for the 
\citet{land} standard stars (panel {\em (a)}: V; panel {\em (b)}: I). 
The open circles represent
standard stars observed during the 1997 January 2 night, while bold face open
triangles represent the standard stars observed during the  1997 January 3 
night. The error bars are formal errors as provided by the IRAF/PHOT package.
The calibration relations are represented by the straight lines; the
corresponding equations and the RMS of the linear fit are reported in the 
insets.
}
\end{figure*} 

The relative photometry has been carried
out with the PSF-fitting code DoPHOT 
\citep{dophot}, running on a Compaq/Alpha station at the Bologna Observatory.
A quadratic polynomial has been adopted to
model the spatial variations of the PSF. Since the code provides a 
classification of the sources, after each application we retained only the
sources classified as {\em bona fide} stars (types 1, 3 and 7).

The relative photometry catalogues from different exposure frames were reported
to a common relative system for each field (INT and OUT) and merged into a 
single final catalogue. 
In particular, since typical intermediate exposures frames sampled
simultaneously the HB and the TO region we carefully checked that the merged
catalogues accurately reproduce the morphology of the CMDs from intermediate
exposures. 
The INT and OUT catalogues were reported to the OUT relative system with the
same accuracy, and the aperture corrections were determined for suitable bright
and uncrowded stars in each of the OUT fields using IRAF/PHOT.
Thus, the final instrumental CMDs are free from any spurious distortion.

Many standard stars taken from the list by \citet{land} have been observed during
both nights to provide a transformation to the standard Johnson-Cousins system.
In Fig.~1 the calibrating color equations are shown. In the present context it
is important to note that (1) the whole color range of the final CMDs is covered
by the observed standards: in the instrumental color index the extreme blue HB 
(BHB) stars have $v-i\sim -1.2$, the reddest RGB tip stars $v-i\sim 0.8$ and the
whole MS and SGB regions are contained between  $v-i\sim -0.6$ and 
$v-i\sim 0$; and (2) the photometry of both nights is tied to the same color
equation. Thus, even if the absolute calibration is not correct {\em the
differential comparisons between the CMDs obtained during this
observational run are fully self-consistent over the whole range of magnitudes
and colors}.
We checked our calibrated data with the photometry by \citet{ros00a} and the
results of this comparison are shown in Fig.~2\footnote{The Rosenberg et al. 
catalogues have been kindly provided in electronic form by G. Piotto.}.

\begin{figure*}[ht]
\figurenum{2}
\centerline{\psfig{figure=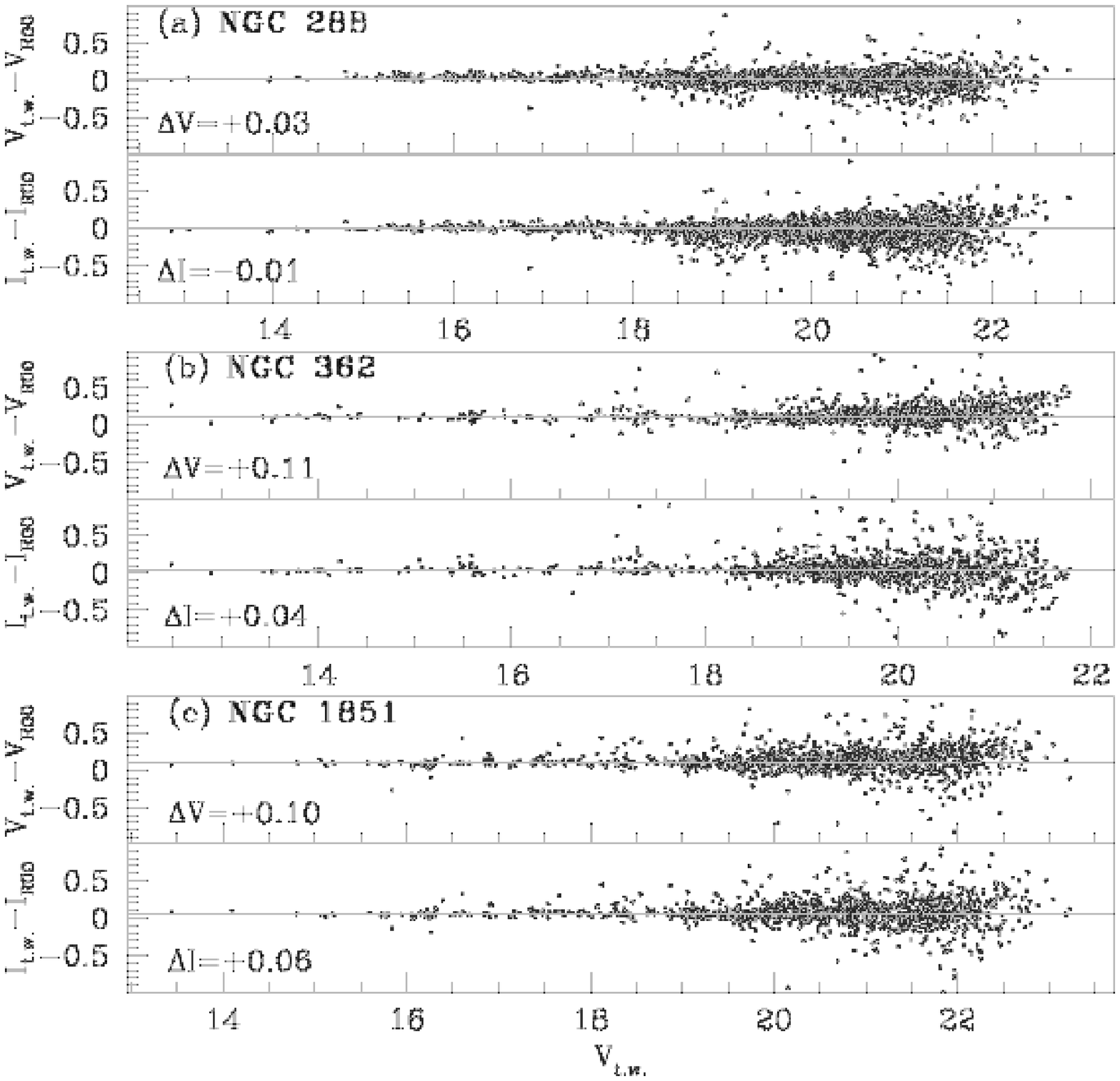}}
\caption{Comparison of the photometry from this work (t.w.) and that of
 \citet{ros00a} (R00). Panel {\em (a)}: NGC~288, panel {\em (b)}: NGC~362,
 panel {\em (c)}: NGC~1851.}
\end{figure*}

While there is good agreement for NGC~288, significant zero-point shifts are 
detected for
NGC~362 and NGC~1851. The presence of thin cirrus during the observations could
be responsible for the small difference with respect to \citet{ros00a}. Thus in
the following we will adopt their absolute calibration. On the other hand, we
emphasize that for the bridge test the absolute calibration is unimportant, 
so this is not a concern. 
 Apart from the quoted shifts the agreement with 
Rosenberg et al. is excellent and the linearity of the relative photometry is
confirmed.

\subsection{Selection of the samples and CMDs}

In order to obtain the best suited samples for the test we selected the stars in
each catalogue according to the following criteria:

\begin{enumerate}

\item All the stars with photometric errors in $V$ or $I$ larger than 0.1~mag have
been excluded from the samples;

\item All the stars of NGC~1851 with $V \ge 18.5$~mag and with a distance from the
cluster center $r<600$~px have been excluded from the sample. All the stars of 
NGC~362 with $V \ge 17.5$~mag and with a distance from the cluster center 
$r<600$~px were also excluded. In this way we avoid confusion in the
SGB-MSTO region of the CMDs due to uncertainties in the photometry of faint stars
in the most crowded part of the fields. In the NGC~288 images the crowding is
moderate everywhere and such selection was unnecessary.

\end{enumerate}

The final selected samples provide very well populated evolved sequences
(collecting stars from the whole observed fields) and very clean SGB-MSTO 
regions populated by relatively uncrowded stars. The catalogues of the selected 
samples are presented in Tables 1, 2 and 3 for NGC~288, NGC~1851 and NGC~362,
respectively. As a fast guidance for the reader we present also the average
photometric errors as a function of $V$ magnitude in Table 4. Note that these
are formal errors as provided by the PSF fitting code, thus taking into account
only the uncertainties associated with the actual $S/N$ ratio of the observed
stars and with the fit process \cite[see][for a discussion]{w92}.

The CMDs from the final samples
are shown in Fig.~3 (panel {\em (a)}: NGC~288, 10990 stars; 
panel {\em (b)}: NGC~1851, 9118 stars; 
panel {\em (c)}: NGC~362, 7499 stars). The encircled dots
in Fig.~3 are the RR Lyrae variables we identified in our samples from
the lists found in the literature: \citet{kal96} for NGC~288; \citet{w98} for
NGC~1851, and the \citet{sh73} catalogue for NGC~362, in the version updated to
1996 by C. Clement and kindly made available in electronic form by the same
author\footnote{See also \citet{clem}; the electronic catalogue can be 
retrieved from: {\tt http://www.astro.utoronto.ca/$\sim$cclement/read.html}.}.  
The open squares in the CMDs of NGC~288 and NGC~1851 are 
the probable bona-fide extreme HB star EHB 1 identified by \citet{bm99} in NGC
288, and two ultraviolet sources identified by the UIT satellite (UIT-31 and
UIT-44) in NGC~1851 \citep{parise}. 
We also cross correlated our catalogue of NGC~288 with the
catalogue of proper motions by \citet{guo}, which partially covers the cluster.
Only one obvious non-member star was identified (GUO 4110), that is not shown in
Fig.~3 because it lies outside the limit of the plot.
NGC~362 is located in the foreground of the Small Magellanic Cloud (SMC) halo,
and a population of  SMC main-sequence  stars and giants can be seen blueward
and redward, respectively, of the NGC~362 main-sequence in Figure~3.  
However,  the SMC contamination is evidently quite small for  $V < 17.5$~mag 
\cite[cf. the NGC~362 background field CMD of][]{l98} 
and thus insignificant for the present purposes.  
We identified three possible foreground stars in the HB
locus of NGC~362 using the astrometric catalog  of \citet{t92}. These
foreground candidates  have both small proper motion errors and low probability
of cluster membership.       
One of these foreground candidates (T245) is
identified along the blue HB in Fig.~3 (open triangle), 
while the other two (T134, T324) are
located near the clump of red HB stars. In Fig.~3 are also indicated two blue 
HB stars (open squares) in NGC~362 which were identified as spectroscopic 
cluster members by \citet{m00} (their MJ 6558 and MJ 8241). 
While the blue HB of NGC~362 is
quite sparse, it does allow a direct comparison with the HB of NGC~288, and can
provide a check on our bridge match.

\begin{figure*}[ht]
\figurenum{3}
\centerline{\psfig{figure=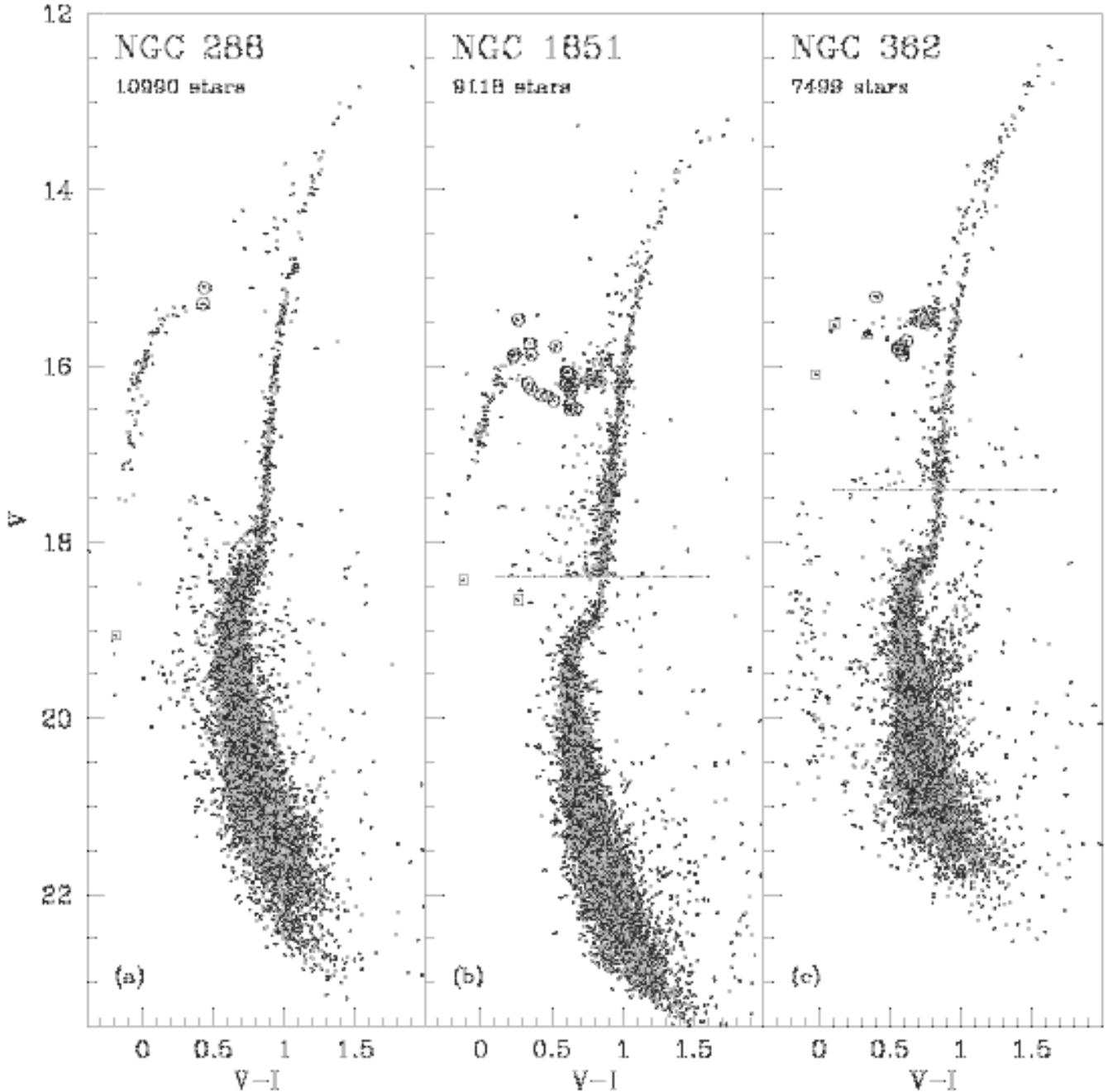}}
\caption{Final CMDs adopted for the bridge test (panel {\em (a)}: NGC~288, 
panel {\em (b)}: NGC~1851; panel {\em (c)}: NGC~362). 
The open circles represent identified RR
Lyrae variables, squares represents previously identified 
UV-bright stars, triangles represent
probable non members (see text). The long dashed horizontal segments in the 
CMDs of NGC~1851 and NGC~362 mark the magnitude limit between the samples
taken from the whole observed fields (bright stars) and those selected from the
less crowded external regions ($r>600$~px; faint stars). }
\end{figure*} 

All the relevant features of the CMDs are very clean and well defined and they
do not show any particular new feature with respect to previous studies,
thus we do not discuss them in detail. Some evidence emerging from the CMD of 
NGC~362 deserves just a brief comment. First, as mentioned earlier, the obvious 
plumes of stars between
$V\sim 17$ and $V\sim 21$ at $V-I\sim 0$ and $V-I\sim 1.1$ are,
respectively, the MS and the RGB of the Small Magellanic Cloud that lie in the
background of this cluster. Second, the asymptotic giant branch 
(AGB) sequence of NGC~362, at $13.5 < $V$ <
15$ and $V-I\sim 1$, is particularly tight and well defined with respect to
that of the other two clusters. This can be due to the very clumpy nature of the
red HB of NGC~362: the evolutionary paths of most stars from the HB along the AGB
are very similar for all the stars of NGC~362 since their initial conditions on
the ZAHB are very similar. On the other hand the AGB of NGC~1851 and NGC 288 
should have a sizeable contribution from stars that were located at very 
different colors along the ZAHB. 

\subsection{Is NGC~1851 a Genuine Globular Cluster?}

Before proceeding any further it is important to address a point that may 
undermine the very basis of the bridge test. It has in fact been suggested
\citep{lee99,yll} that GCs with multimodal HB morphology
(including NGC~1851) are not {\em single age - single metallicity} objects as 
any classical globular, but do host populations of different metallicity and/or 
age that would be responsible for the anomalous HBs. 
The hypothesis is not new and
reappears in the literature from time to time under different forms
\citep{a90,syd96}, but no observational evidence supporting this view has been
found yet \citep{w92,cat97}. It is interesting to note that in the only proven 
case of a globular with multiple population, i.e. $\omega$~Centauri
\citep{lee99b,panc}, the metallicity spread was immediately recognized as soon
as the first modern CMD was assembled \citep{dw67}. 

The case of NGC~1851 has been studied in detail by \citet{w92}, who, from the
observed width of the RGB and MS, concluded that any possible metallicity
dispersion in  this cluster is $\sim 0.1$ dex in ${\rm [Fe/H]}$, in excellent
agreement with the analogous results by \citet{da90}. We measured the color
width of the RGB at the HB level (in the range $V_{\rm HB}\pm 0.2$) in order to 
obtain a $\sigma {\rm [Fe/H]}$ from $\sigma (V-I)$ through the calibration of the 
$(V-I)_{\rm 0,g}$ parameter [see \citet{sav98}, for definition, calibration and 
references]. We find $\sigma (V-I)=0.016, 0.024, 0.026$~mag for NGC~362, NGC~288
and NGC~1851, respectively, in good agreement with the above results. Adopting
the calibration by \citet{sav98} the maximum metallicity dispersion 
of NGC~1851 is $\sigma {\rm [Fe/H]} < 0.15$~dex, neglecting all observational 
sources
of scatter along the RGB. In this context it is more important to note that the
RGB width of the three considered clusters, observed under similar conditions, 
is very similar, indicating that no significant intrinsic difference in the 
color distribution of the RGB is observed.   

An analogous test concerning possible age spreads has been performed by studying
the dispersion in $V$ around the ridge line in the most horizontal part of the 
SGB ($0.7<(V-I)<0.8$), that we will adopt in \S4.4 as an age diagnostic.
It turns out that $\sigma_V=0.082, 0.099, 0.080$~mag 
for NGC~362, NGC~288, and NGC~1851, 
respectively. Again, the result is {\em very similar} for the three clusters
suggesting that if an intrinsic age spread were present in NGC~1851, so it 
would also in the cases of NGC~288 and NGC~362.
Neglecting observational errors and assuming that all the observed $\sigma_V$ is
due to an intrinsic spread in age, this is constrained to 
$\sigma ({\rm age})\le 0.8$~Gyr, 
according to the age scale illustrated in \S4.4 below.

It can be concluded that the existing data constrain any possible age and
metallicity spread in NGC~1851 to small amounts, so that the explanation 
of the anomalous HB morphology in these terms can be excluded. On the other hand,
\citet{sav98} found that the radial distribution of BHB stars is significantly
different from that of RGB and SGB stars in the outer region of the cluster
($r>50\,r_c \sim 100$~arcsec). Such evidence can be more easily reconciled with
scenarios in which dynamical processes favor the production of BHB stars, 
in agreement with the results by \citet{fb93} and \citet{buo97}, 
than with the hypothesis of a stellar system with multiple populations. 

\section{The Bridge Test}

Important tools for the actual application of the bridge test are the ridge 
lines of the clusters on the CMDs. 
We derived the ridge lines by averaging and 2-sigma clipping 
on boxes of different sizes, depending on the density of stars and on the
required resolution, in different regions of the CMDs 
\cite[an approach similar to that adopted by][]{fer99}. 
The ridge lines of the MS and RGB sequences 
are presented in Table~5, while the HB ridge lines are reported in Table~6.
In Fig.~4 a zoomed view of the MSTO region is presented,
to allow a direct comparison of the
adopted fiducials with the data for this crucial part of the CMDs.
 
\begin{figure*}[ht]
\figurenum{4}
\centerline{\psfig{figure=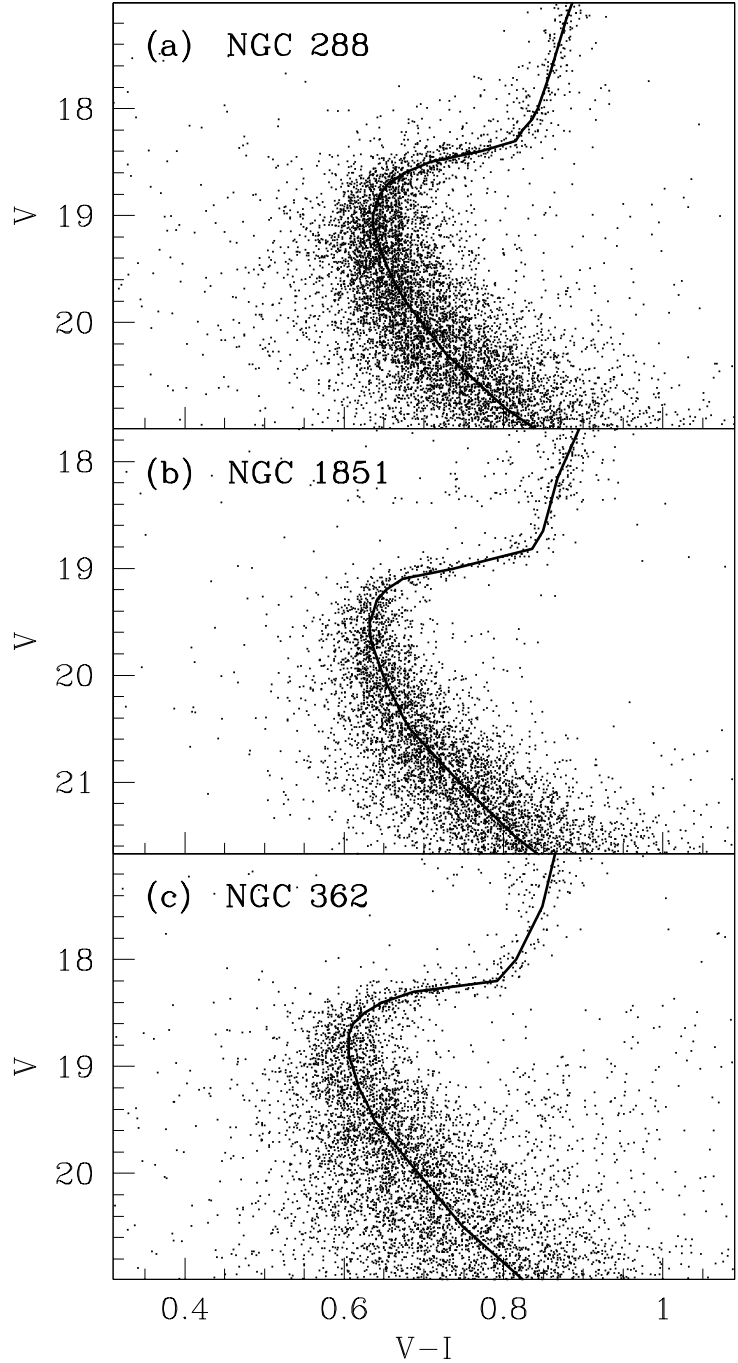}}
\caption{Comparison between the data and the adopted ridge lines in the region
of the CMD surrounding the MSTO point.Panel {\em (a)}: NGC~288, 
panel {\em (b)}: NGC~1851; panel {\em (c)}: NGC~362. Note, in particular the
excellent fit of the SGB sequence, that will be the key observable in the final
measure of the age difference (\S4.4).}
\end{figure*} 
 
\subsection{Matching HBs}

The bridge test has been conceived essentially as an age test. Therefore, 
to remove the undesired effects of distance and reddening, it 
seems much more advisable to find the best match for the less age-dependent
features of the CMDs (HB luminosity level, RGB) and to
check the consequent agreement or disagreement between the more age-sensitive
features (SGB, MSTO). Furthermore, HB stars are less plagued by crowding effects
and the same morphology of the features provides a much firmer benchmark for
matching with respect to the nearly vertical MSTO region. For these reasons we
adopt the matching of the HBs as the preferential version of the bridge test
and we try the approach of SVB96 only as a check. The final results are
independent of the adopted strategy.

The match between the BHBs of NGC~288 and NGC~1851 is presented in Fig.~5. The
ridge line of NGC~1851 (heavy lines in both panels) is superposed to the
observed HB of NGC~288 (filled circles). An excellent match is found if the 
shifts $\Delta V = +0.66$~mag and $\Delta (V-I) = +0.015$~mag are applied 
to report NGC~288 upon NGC~1851. 
Panel {\em (a)} of Fig.~5 shows the effects of the adoption of
different shifts in $V$ by reporting copies of the NGC~1851 ridge line shifted by 
$\pm 0.05, 0.10$ mag in $V$ (thin lines) with respect to the true ridge line. It
can be readily appreciated that a difference of $\pm 0.05$ mag with respect to
the adopted shift would clearly produce a poor match between the HBs. Since the
same compatibility range is obtained if the NGC~1851 data points are compared
with the ridge line of NGC~288, we add in quadrature the two terms to take into
account the uncertainties of both the ridge lines, so obtaining a 
global---conservative---compatibility range of $\pm 0.07$~mag. 
The
effects of changes in the adopted color shift are reported in the panel {\em (b)} 
of Fig.~5 in a strictly analogous way by reporting copies of the NGC~1851 ridge
line shifted by $\pm 0.01, 0.02$ mag in $V-I$. A difference of $\pm 0.01$~mag 
with respect to the adopted color shift is---at most---marginally acceptable. 

We followed the approach presented in Fig.~5 since the adopted shifts are the
basis of the bridge test, thus their reliability and the associated
uncertainties have to be firmly assessed. As a final choice we conservatively
adopt the compatibility ranges found in Fig.~5 as the uncertainty in the shifts.
For NGC~288: $\Delta V = +0.66 \pm 0.07$~mag 
and $\Delta (V-I) = +0.015 \pm 0.01$~mag.

\citet{vdb00} and \citet{gru99} (as well as the analysis presented in
Paper~II) suggest that the brightest and reddest stars 
in the HB of NGC~288 are somehow anomalous and/or significantly evolved, 
therefore
they can cause misleading matches when compared with HB stars of other clusters.
We emphasize that our match between the HBs of NGC~288 and NGC~1851 is primarily
based on the match of the bluer part of the distributions. This is a rather
inescapable choice, since these are the best populated and tightest parts of 
the HBs. In particular the tight sequence of the HB of NGC~288 in the range 
$0.0\le (V-I) \le 0.07$ (see Fig.~5 and 6) is the key feature for the adopted 
match.
We note that fixing this feature an excellent overall fit is obtained, thus the
presented match is unlikely to be affected by the possible
anomalous status of the stars discussed by the quoted authors. The point is
clearly illustrated in Fig.~6 where we reported the BHB stars of NGC~1851
(crosses) and of NGC~288 (open circles) after the application of the obtained
shifts. The region of the BHB that provides the basis for the determination of 
the shift is enclosed in a square. The adopted ridge line of the BHB of NGC~1851
is also overplotted to allow a direct comparison between the data and the
fiducial. The red part of the BHB ridge line is reported (and will be reported
in the following plots) as a dotted curve to put in evidence that {\em (a)} 
this is the most uncertain part of the NGC~1851 ridge line and, 
{\em (b)} {\em it has not been taken into account} in our shift determination.
It can be readily appreciated that we made all efforts to avoid any possible
ambiguity in the determination of the shift between NGC~288 and NGC~1851.
Nevertheless, we want to stress again that the match of the blue HBs is the 
most uncertain passage of the whole bridge test, from an observational point 
of view.

\begin{figure*}[ht]
\figurenum{5}
\centerline{\psfig{figure=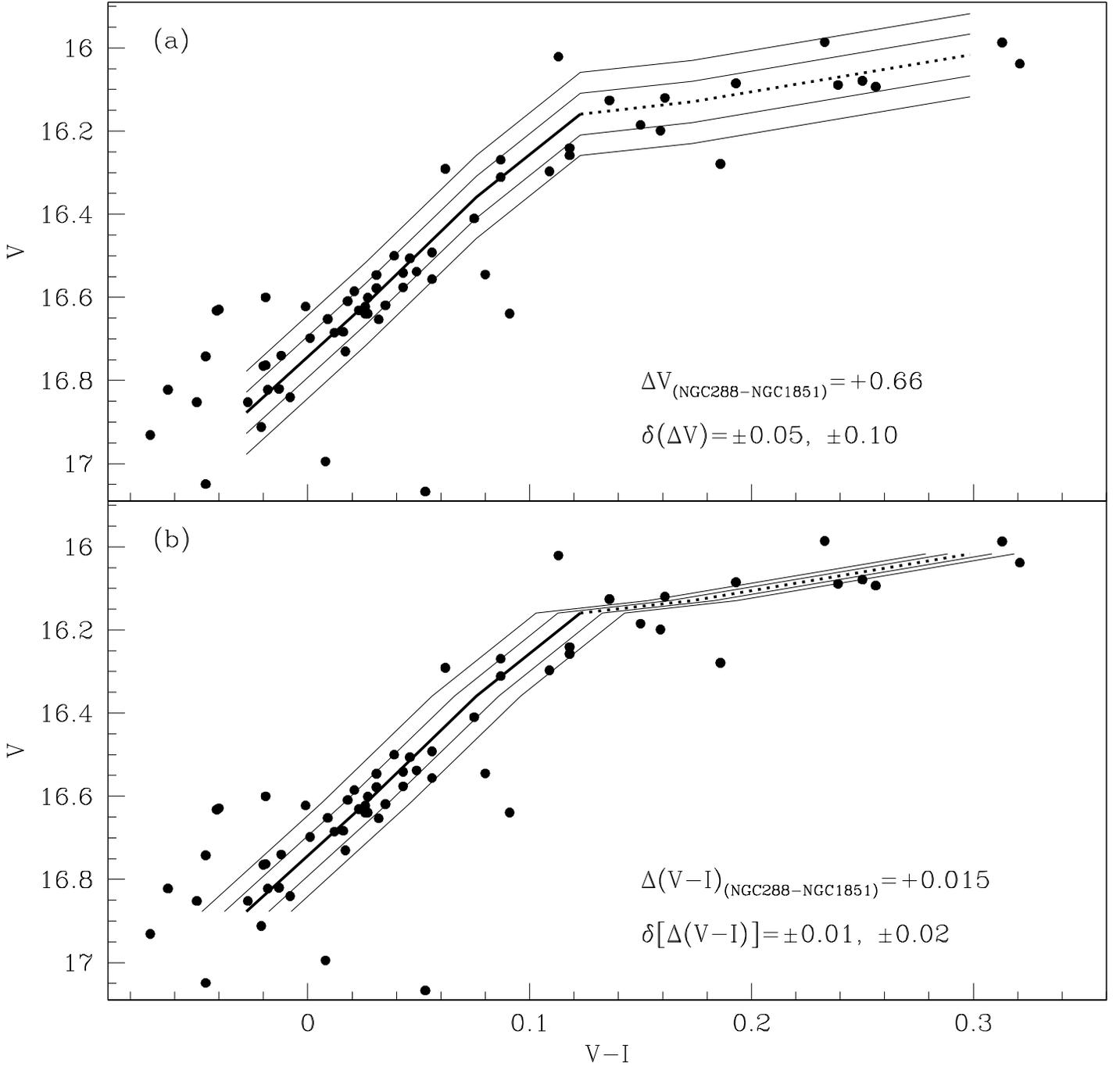}}
\caption{Matching the BHB of NGC~288 (filled circles) to the BHB ridge line of 
NGC~1851 (heavy line; the reddest part of the ridge line is represented as
a dotted line to recall that it is the most uncertain part of the adopted
fiducial and, above all, that we {\em do not use this part} in the
determination of the shift; see Fig.~matchhb). 
Panel {\em (a)}: the effect of different assumptions for 
the V shift is shown by reporting the NGC~1851 ridge lines shifted by 
$\pm 0.05, 0.10$~mag with respect to the adopted best fit shift 
$\Delta V=+0.66$~mag.
Panel {\em (b)}: the effect of different assumptions for the $V-I$ shift is 
shown by reporting the NGC~1851 ridge lines shifted by $\pm 0.01, 0.02$~mag 
in $V-I$ with respect to the adopted best fit shift $\Delta (V-I) = +0.015$~mag. 
In both cases it is evident that the heavy lines provide the best match and
that differences as small as $\pm 0.05$~mag in $V$ and/or $0.01$~mag in $V-I$ with
respect to this line provides a much poorer fit to the data.}
\end{figure*} 

\begin{figure*}[ht]
\figurenum{6}
\centerline{\psfig{figure=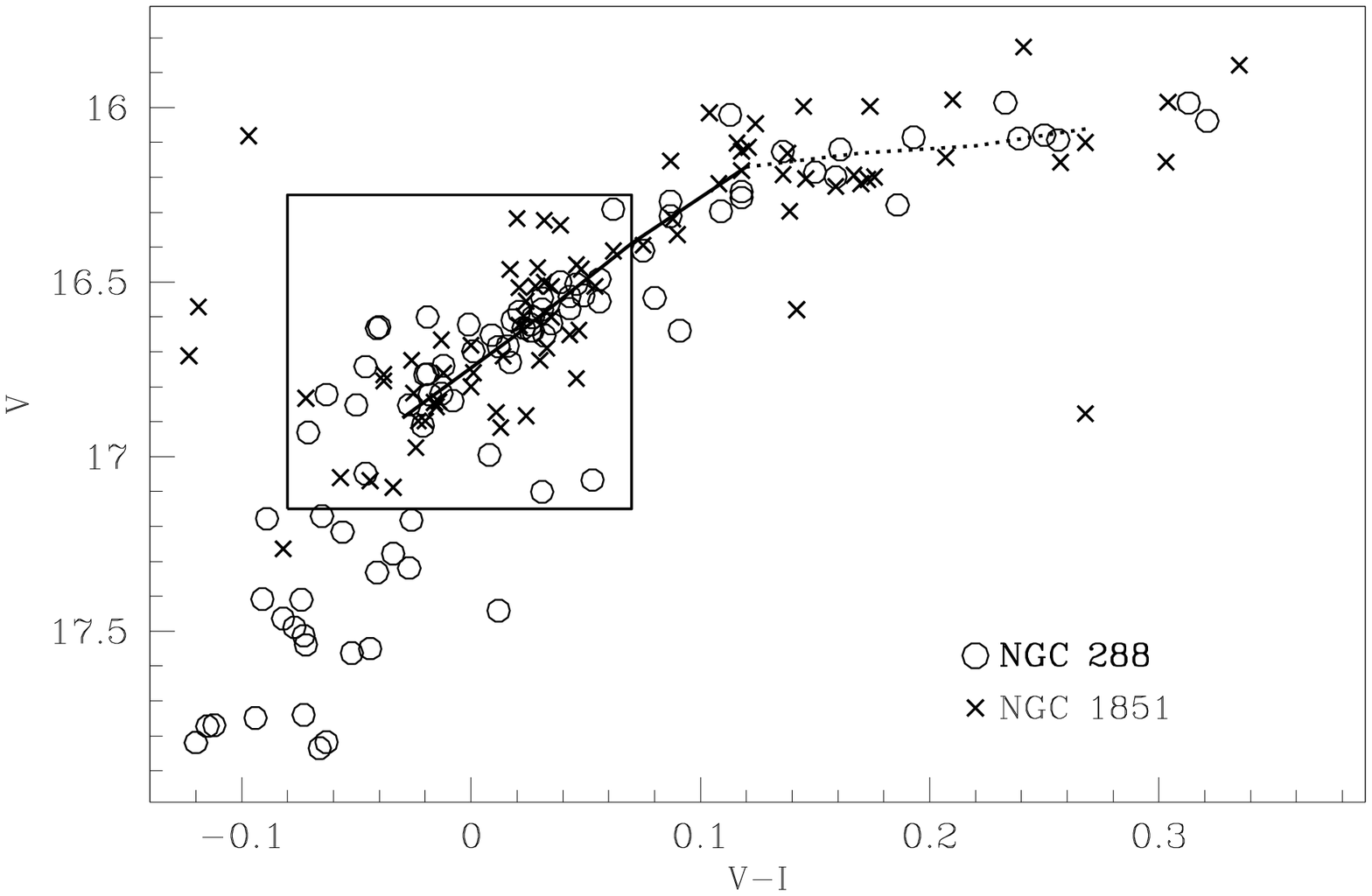}}
\caption{Superposition of the non-variable BHB stars of NGC~1851 (crosses;
the stars of NGC~1851 with $V\ge 17.5$ have not been plotted in this panel) 
and of NGC~288, after the application of the obtained shift. The square enclose
the blue prt of the BHB we use to determine the shift. Whith this approach
we avoided to base the determination of the shift on the reddest BHB stars
whose evolutionary status is uncertain. The BHB ridge line of NGC~1851 is also
reported to allow the reader to campare it directly with the data. The reddest 
part of the fiducial is reported (and will be reported in all the following
plots) as a dotted line to recall that it has not been considered in the
determination of the shift between NGC~288 and NGC~1851. Note, however, the
good overall fit}
\end{figure*} 

The match between the well populated and clumped red HB (RHB) 
of NGC~362 and NGC~1851
is a much simpler task.
Panel {\em (a)} of Fig.~7 shows the excellent match between the histograms of
the RHB of NGC~1851 (dashed heavy line) and of the RHB of NGC~362 (thin
continuous line) that is obtained applying a shift of $\Delta V = 0.665$ mag
to NGC~362. 
The clear peaks in the distributions offer a robust
reference to derive the optimal shift. In panel {\em (b)} of Fig.~7 the
cumulative distributions of RHB stars in $V$ are compared by adopting slightly 
different shifts. 
The thin solid lines represent the distributions of the RHB of NGC
362 after the application of the shifts $\Delta V = +0.650, 0.665, 0.680$~mag, 
from
left to right respectively. It is evident that a difference of $\pm 0.015$~mag
with respect to the adopted $V$ shift would provide a much poorer fit between the
two distributions. The color shift has been found by comparing also the RGBs and
the final adopted shifts to report NGC~362 upon NGC~1851 are 
$\Delta V = +0.665 \pm 0.015$~mag and $\Delta (V-I) = 0.03 \pm 0.01$~mag.   
The adopted shifts are in agreement, to within the uncertainties, with the
current estimates of distance moduli and reddening differences among the
considered clusters \cite[see e.g.][]{fer99}. The final difference in the
apparent distance moduli between NGC~362 and NGC~288 that is implicitly obtained
with the derived shifts is $\Delta \mu = -0.005 \pm 0.087$, where the errorbar
should be considered as a compatibility range, which takes into account all the
uncertainties.

\begin{figure*}[ht]
\figurenum{7}
\centerline{\psfig{figure=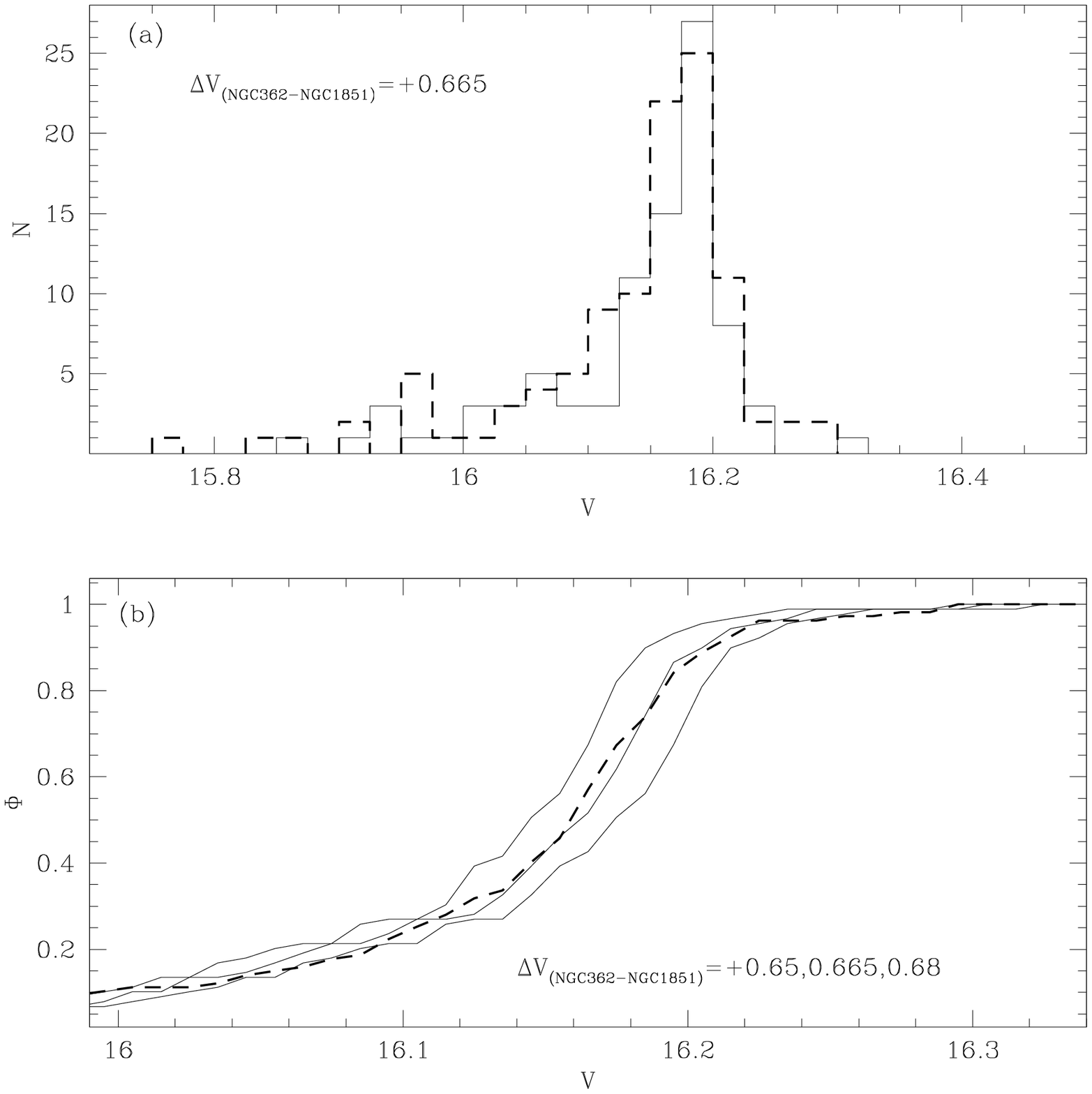}}
\caption{Matching the RHB of NGC~362 (thin solid line) to the RHB of 
NGC~1851 (heavy dashed line). 
Panel {\em (a)}: The $V$ histogram of the two distribution after having applied 
a shift of $\Delta V= +0.665$ mag to the NGC~362 data. 
Note the excellent match of the peaks of the histograms. 
Panel {\em (b)}: Comparison between the cumulative
distributions after the application of different shifts to the NGC~362 data as
reported in the lower right corner of the panel. A difference of $\pm 0.015$ mag
with respect to our best fit value is already inconsistent with the data.}
\end{figure*}

The overall match among the ridge lines after application of the above
shifts can be judged from Fig.~8. The agreement is as good as possible for the
HB (obviously) but is also very good
for the whole RGB and lower MS. The only
significant difference appears in the MSTO region of the CMD, which 
anticipates the main result that will be discussed below: the MSTO and SGB
sequences of NGC~288 are fainter and redder than those of NGC~362 and NGC~1851, 
while the latter two clusters seem similar in this regard.

\begin{figure*}[ht]
\figurenum{8}
\centerline{\psfig{figure=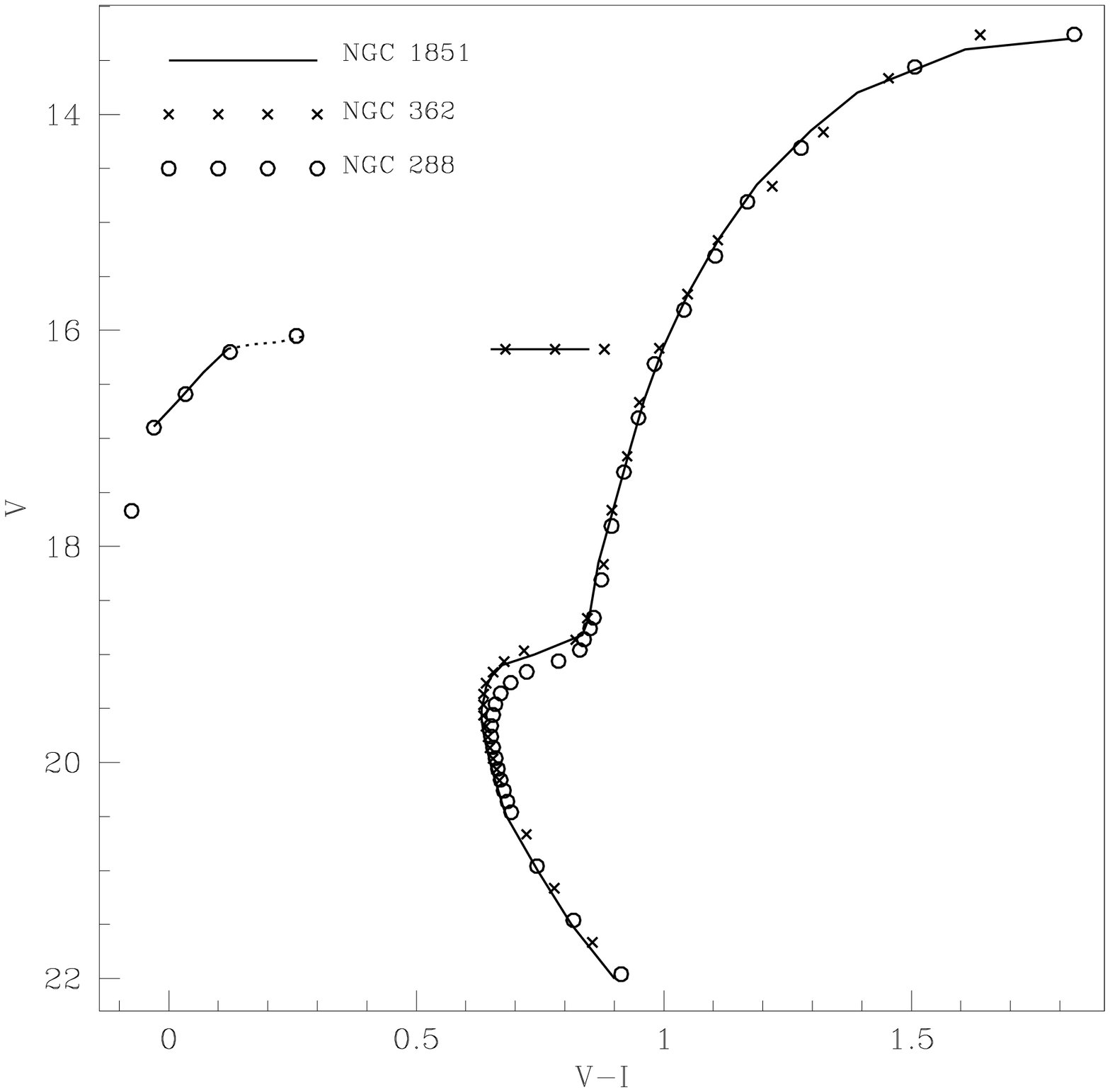}}
\caption{Overall match of the ridge lines of the three clusters after the
application of the derived shifts to report the HB of NGC~362 and NGC~288 upon 
that of NGC~1851. Solid line: NGC~1851; crosses: NGC~362; open circles:
NGC~288.}
\end{figure*}

In Fig.~9 a close view of the two crucial regions of the CMD is provided to make
clear the effects of the adopted matching. In panel {\em (a)} it can be
appreciated the fine agreement between the NGC~288 (open dots) and NGC~362
(crosses) data points and the ridge line of NGC~1851 in the HB region of the
CMD. In panel {\em (b)} the shifted ridge lines of NGC~288 (heavy solid 
line and  open circles) and of NGC~362 (heavy dashed line and crosses) are 
reported. It
is readily evident that the MSTO and the SGB of NGC~362 are significantly
brighter and bluer than those of NGC~288. The differences are larger than the 
maximum errors in the applied shifts. 

All other parameters being fixed, we are led to the conclusion that 
{\em NGC~362 is younger than NGC~288}. 
We will quantify such an age difference in \S4.3 and 4.4. 

\begin{figure*}[ht]
\figurenum{9}
\centerline{\psfig{figure=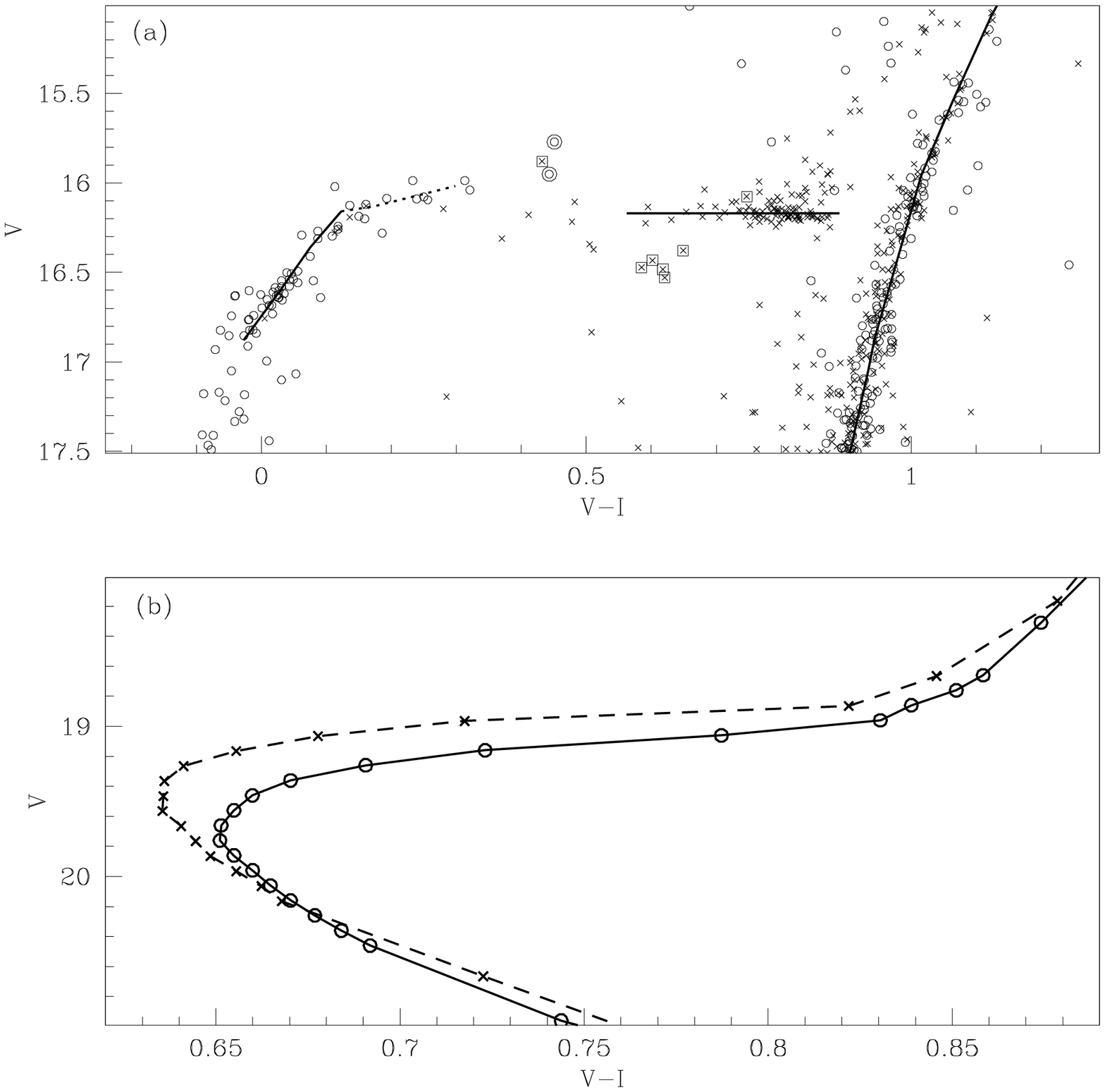}}
\caption{Panel {\em (a)}: global match of the HBs on the CMD. The open dots 
represent the NGC~288 data, 
the encircled symbols denoting the RR Lyrae variables; the crosses
are the NGC~362 data, those enclosed in an open square being 
the identified RR Lyrae variables.
The solid line is the ridge line of NGC~1851. 
Panel {\em (b)}: close up view of the
MSTO region of the CMD. The ridge lines of NGC~288 (heavy solid line and
open circles) and NGC~362 (heavy dashed line and crosses) are reported, once
shifted according to the amounts needed to match the HBs.}
\end{figure*}
 
\subsection{Matching MSTOs}

The result of the bridge test in the version introduced by SVB96 (i.e.,
matching MSTOs and checking the agreement of the HBs) is shown in Fig.~10, which
is analogous to Fig.~9. In panel {\em (b)} the ridge lines of NGC~288 (open
circles) and NGC~362 (crosses) have been shifted to provide the best match to
the ridge line
of NGC~1851 (heavy solid line) in the MSTO and SGB region. 
The adopted shifts are $\Delta V = +0.540$ and $\Delta (V-I) = -0.005$ for
NGC~288, and  $\Delta V = +0.672$ and $\Delta (V-I) = +0.026$ for NGC~362. 
The overall
fit is excellent, nevertheless a mismatch in the color of the base of the RGB of
NGC~288 and NGC~362 is evident and turned out to be unavoidable (and has been
noted also by \citet{vdb00}). Note that the observed mismatch is consistent with
the existence of an age difference between the two clusters, NGC~288 resulting
the older one (see below).
The same shifts have been applied to the NGC~288 and NGC~362 data and the
corresponding CMD of the HB region is shown in panel {\em (a)} of Fig.~10 and
compared to the ridge line of NGC~1851 (the symbols are the same as in 
panel {\em (a)} of Fig.~9). The mismatch between the data and the ridge line is well
beyond the compatibility range, that is reported as a couple of thin lines
paralleling the HB of NGC~1851. Note that the adopted shift introduces an 
obvious mismatch also among the upper RGB sequences. 

Therefore there is no room for a simultaneous superposition of the HBs and the
MSTOs of the three considered clusters. {\em We conclude that 
this version of the bridge test also implies 
a significant age difference between NGC~362 and NGC~288, 
the former being the younger cluster.}

\begin{figure*}[ht]
\figurenum{10}
\centerline{\psfig{figure=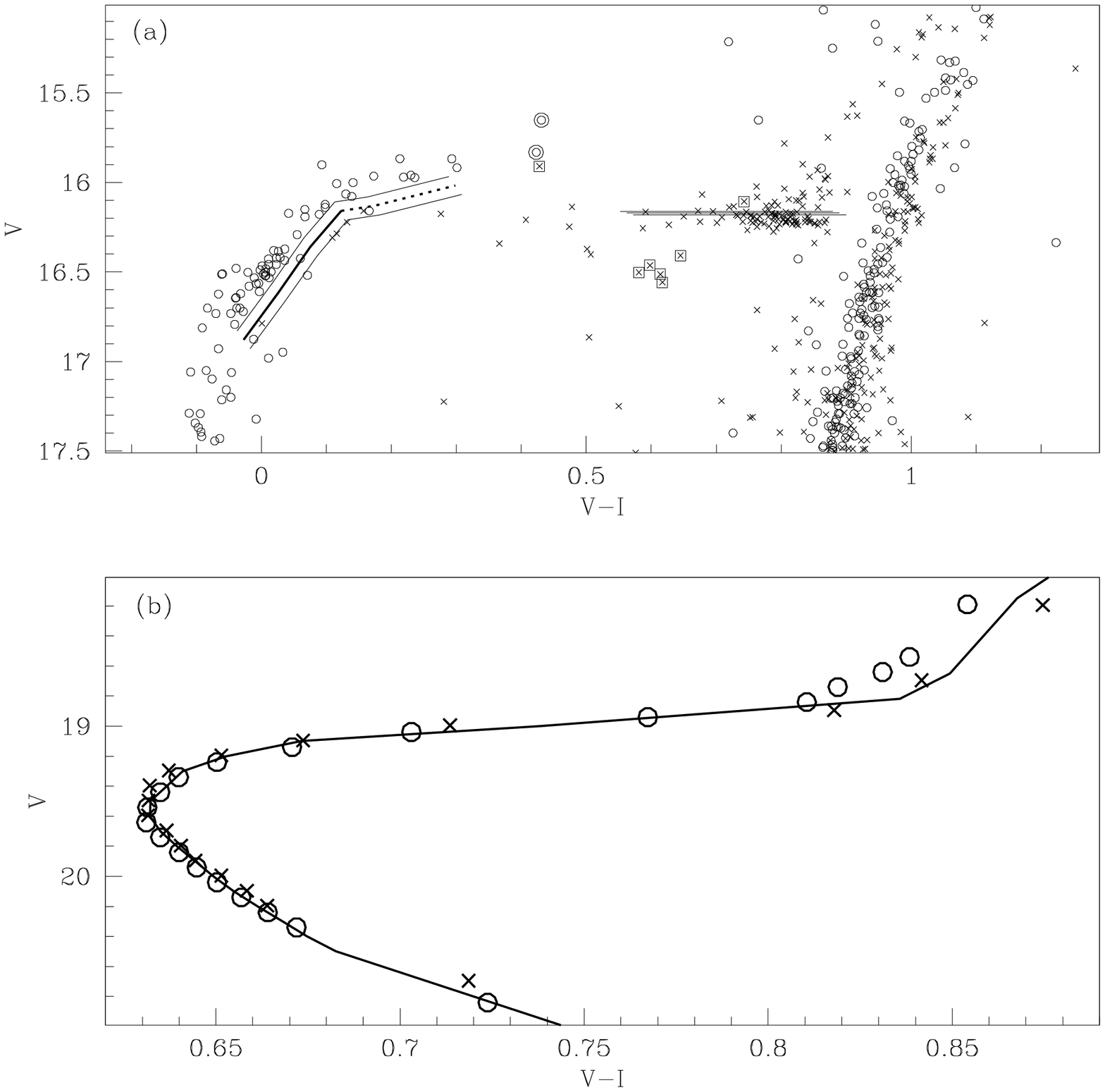}}
\caption{Panel {\em (b)}: the ridge lines of NGC~288 (open
dots) and NGC~362 (crosses) shifted to provide the best fit to the NGC~1851 
ridge line
(heavy solid line) in the MSTO and SGB region. Panel {\em (a)}:
the HBs of NGC~288 (open dots) and NGC~362 (crosses) after the application of
the shift derived from the match shown in panel {\em (b)}, 
compared to the ridge line of NGC~1851 (heavy line). 
The thin lines paralleling the ridge line bracket
the range of maximum acceptable difference in the shifts.
Note that a small mismatch ($\le 0.05$ mag) is apparent also between the RHB 
of NGC~362 and NGC~1851, suggesting that the former may be slightly younger 
than the latter, as shortly discussed in \S4.3. }
\end{figure*}

\subsection{Differential age parameters}

As a first method to quantify the detected age differences we use the
same differential age parameters adopted by \citet{ros99}, namely:

\begin{itemize}

\item{} The {\em horizontal} parameter $\delta (V-I)_{@2.5}$, i.e. 
the difference in color between the MSTO point and the point 2.5~mag brighter 
in $V$ on the base of the RGB;

\item{} The {\em vertical} parameter $\Delta V_{\rm TO}^{\rm HB}$, i.e. the difference
in magnitude between the HB level and the MSTO point.

\end{itemize}

We refer the reader to 
SVB96 and \citet{buo98a} for a thorough discussion about these
parameters. The close similarity in metal content
will greatly contribute to reduce the uncertainties in the final age difference 
estimates from both observables, especially for NGC~362 and NGC~288 (cf. \S2.1).

We start with the horizontal parameter, whose application is independent of
the accurate matching obtained for the bridge test. From the
ridge lines of the clusters we obtain 
$\delta (V-I)_{@2.5}({\rm NGC~288})=0.276\pm 0.010$~mag,
$\delta (V-I)_{@2.5}({\rm NGC~1851})=0.306\pm 0.010$~mag, and 
$\delta (V-I)_{@2.5}({\rm NGC~362})=0.301\pm 0.010$~mag, in excellent agreement with 
the
results of \citet{ros99}. From their Figs.~4 and 9, we obtain an age difference
$\Delta {\rm Age}_{\rm NGC~288-NGC~362} = 2.2 \pm 1$~Gyr by averaging over the 
(fully compatible) estimates derived by adopting different sets of models
(\citealt[][hereafter SCL9;]{scl97} \citealt[][hereafter CCDW98;]{c98}
\citealt[][hereafter V2000]{vdbal00}).  
The age difference between NGC~362 and NGC~1851 is found to be null to
within the errors.

Given the optimal match obtained between the HBs of the considered clusters we
can obtain differences in $\Delta V_{\rm TO}^{\rm HB}$ just by subtracting the
observed $V_{\rm TO}$ values once the shifts presented in \S4.1 are applied.
In this way most of the uncertainties associated with the determination of the
HB level are avoided. We obtain 
$\Delta V_{\rm TO}^{\rm HB}({\rm NGC~288})-\Delta V_{\rm TO}^{\rm HB}({\rm NGC~362}) = 
0.23 \pm 0.07$~mag 
and
$\Delta V_{\rm TO}^{\rm HB}({\rm NGC~288})-\Delta V_{\rm TO}^{\rm HB}({\rm NGC~1851}) = 
0.18 \pm 0.07$~mag, just
a few hundredths of a magnitude 
less than found by \citet{ros99}, i.e., $0.26\pm 0.10$ and
$0.23\pm 0.10$, respectively, but still in good agreement. 
The resulting age differences are 
$\Delta {\rm Age}_{\rm NGC~288-NGC~362} = 2.7 \pm 1.0$~Gyr and 
$\Delta {\rm Age}_{\rm NGC~288-NGC~1851} = 2.3 \pm 1.0$~Gyr.  
Also in this case the estimates
from different sets of models are in good agreement, to within the errors. 
The age difference between NGC~362 and NGC~1851 can be considered marginal
(see below).

Therefore, independently of the adopted set of theoretical models and on the
differential age parameter used, {an age difference of $2.4 \pm 1.5$~Gyr is
measured between NGC~288 and NGC~362}, the reported error bar covering the whole
compatibility range spanned by the two independent estimates.
Any derived age difference slightly depends on the absolute age of the oldest
cluster (the age zero point). The above result can be appreciably changed by
this effect only if an age lower than 12~Gyr is assumed for 
NGC~288\footnote{However, if an absolute age of 10~Gyr is assumed for this
cluster the age difference with NGC~362 is still larger than 1.5~Gyr. See also
Paper II.}. The
observed $\Delta V_{\rm TO}^{\rm HB}$ as well as the direct comparison with 
various sets of isochrones suggest instead an absolute age $\ge 13$~Gyr for this
cluster, although this value is subject to possible systematic errors
\cite[see, e.g.,][]{vdb96,dan}. 

While the uncertainty in the 
abundance of NGC~1851 renders its age estimate less reliable, 
our test strongly suggests that NGC~1851 is significantly younger than
NGC~288 and has an age similar to that of NGC~362\footnote{If only the
vertical parameter is considered, we note a marginal indication that NGC~362 may
be slightly younger than NGC~1851. A similar result is obtained also by
\citet{ros99} and it seems confirmed in Fig.~8 and 10. Since NGC~1851 is
not the main target of the present analysis and since the detected difference is
quite small we mantain the conclusion that NGC~362 and NGC~1851 have 
{\em similar} age.}. 
Because of these uncertainties and
since our goal is to study in detail the couple NGC~362/NGC~288 we will
concentrate on these clusters in the remainder of this analysis, as well as 
in Paper~II, leaving NGC~1851 (the ``bridge'' cluster) for occasional reference.

\subsection{A Global Comparison}

In Fig.~11 the ridge lines of NGC~288 (thick solid line) and NGC~362
(thick dashed line), properly shifted according to the results of the matching
of the HBs, are superposed to three different sets of isochrones of the
appropriate metal content\footnote{The V2000 isochrones 
\cite[see also][]{berg01} have been kindly provided 
by D.A. VandenBerg. The SCL97 isochrones have been retrieved from the ORFEO 
database ({\tt http://www.mporzio.astro.it/$\sim$mandrake/orfeo.html}). The CCDW
isochrones have been retrieved from the GIPSY database
({\tt http://www.mporzio.astro.it/$\sim$mkast/GIPSY/homegipsy.html})}. 
The age range covered by each set is $8-18$~Gyr from
left to right. Each set of isochrone has been
reported to the same distance and reddening as the ridge lines by applying the
distance modulus and reddening of NGC~1851, as taken from 
\citet[][from column 8 of their Tab.~2]{fer99}.
Minor adjustments have been made to best fit the nearest isochrone to the
ridge line of NGC~288 (see below). 

In panel {\em (a)} the comparison is
with the SCL97 isochrones with standard helium abundance $Y=0.23$, solar element 
mix and $[{\rm Fe/H}]=-1.0$. According to \citet{scs93} this is the proper set of
isochrones to compare against clusters with $[{\rm Fe/H}]=-1.3$ and a typical halo
$\alpha$-enhancement \cite[e.g.,][]{carn}, as is the case for NGC~288 and 
NGC~362 (cf. \S2.1). 
A simple color shift of $-0.01$~mag in $V-I$, applied to the whole set, 
reports the 13~Gyr isochrone to an excellent superposition with the ridge line 
of NGC~288. The age step between isochrones is 1~Gyr.
In panel {\em (b)} the ridge lines are compared with the CCDW98 set of 
isochrones at $[{\rm Fe/H}]=-1.31$, standard helium and solar element ratios.
The $[{\rm Fe/H}]=-1.0$ isochrones are not present in the CCDW98 set thus we cannot
properly account for $\alpha$-enhancement in this case. The 13 Gyr isochrone was
fitted to the NGC~288 ridge line by shifting the whole set by $-0.004$~mag in 
$V-I$. The comparison with the $Y=0.237$, $[{\rm Fe/H}]=-1.31$, 
$\alpha$-enhanced isochrones by V2000 is shown in panel {\em (c)} of Fig.~11.
A shift of $+0.023$~mag in $V-I$ has been applied to the whole set.  
In this case the age step between the isochrones is 2~Gyr.

\begin{figure*}[ht]
\figurenum{11}
\centerline{\psfig{figure=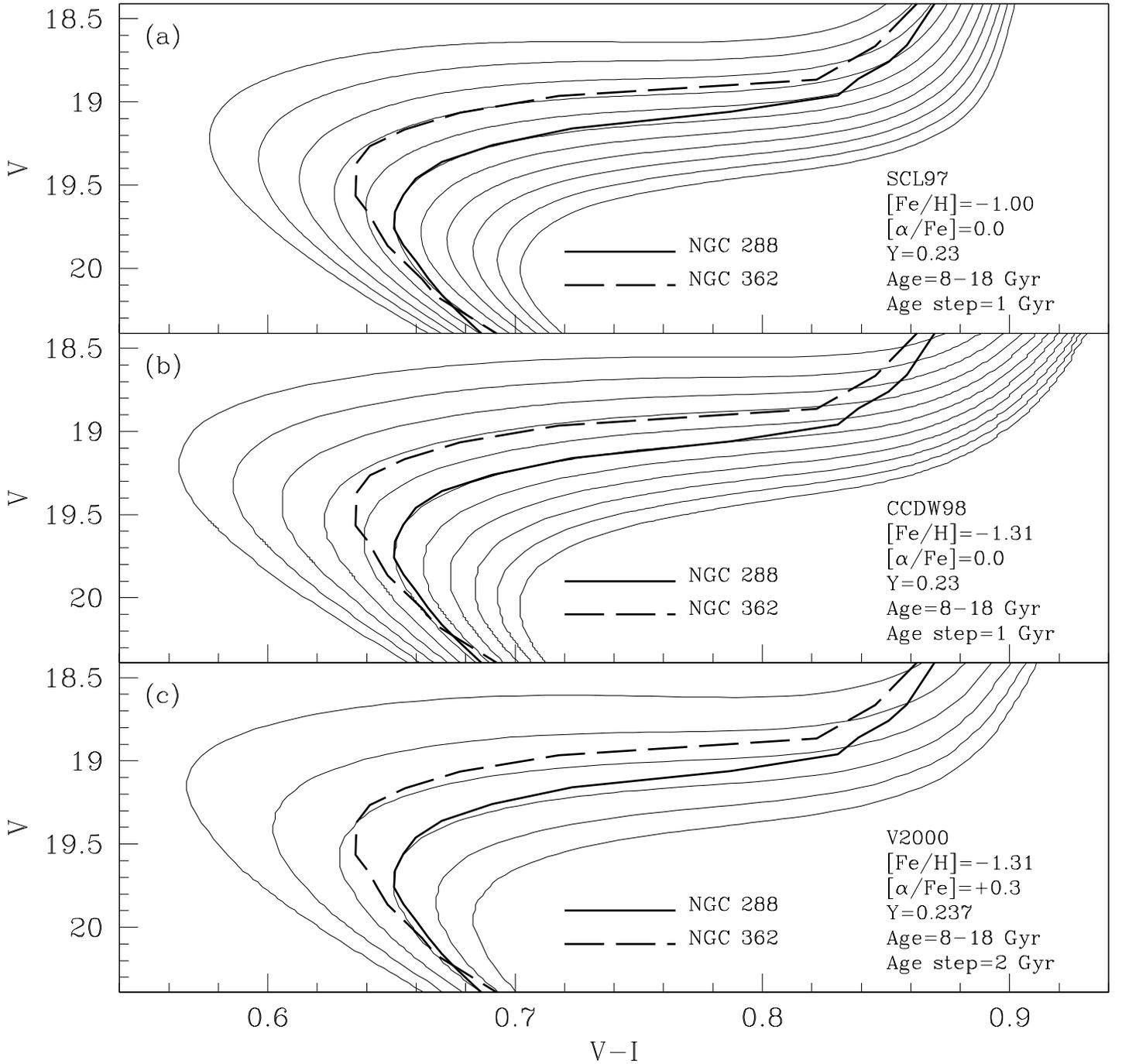}}
\caption{Shifted ridge lines of NGC~288 (thick solid lines) and NGC~362 
(thick dashed lines) are compared to three different sets of isochrones of the
appropriate metal content (thin lines). The details of each adopted
set are reported in the lower right corner of each panel.
Panel {\em (a)}: SCL97 models with $[{\rm Fe/H}]=-1.0$, simulating 
$\alpha$-enhanced isochrones with  $[{\rm Fe/H}]=-1.3$. 
Panel {\em (b)}: CCDW98 models. 
Panel {\em (c)}: $\alpha$-enhanced models by V2000. 
The age step between two adjacent isochrones
is 1~Gyr in panels {\em (a)} and {\em (b)}, and 2~Gyr in panel {\em (c)}. 
In all
cases the ``oldest'' isochrone has an age of 18~Gyr and the ``youngest'' 8~Gyr.
}
\end{figure*}

Despite the differences in the models and in the assumptions, the results of all
the comparisons are very similar. The optimal removal of the effects of distance
and reddening we have obtained with the ``bridge matching'' and the close
similarity in chemical composition allow a very robust approach to the estimate
of the age difference between NGC~288 and NGC~362 from Fig.~11. The nearly
horizontal region of the SGB, between $V-I=0.7$~mag and $V-I=0.8$~mag is a 
well defined (and easy to measure) observational feature that provides a 
natural age scale and that can be 
straightforwardly compared to model predictions. From this comparison an age 
difference of 2~Gyr is clearly detected, independent of the adopted set of
isochrones. It is worth noting that if the luminosity of the MSTO points are
considered, a slightly larger difference is obtained (2.5 to 3~Gyr) in good
agreement with what found with the differential vertical parameter 
$\Delta V_{\rm TO}^{\rm HB}$, in the previous section. 
Given the quoted uncertainties
associated with the {\em measure} of $V_{\rm TO}$ we consider the estimate 
obtained
from the horizontal region of the SGB as more reliable and we adopt it as our
final value, recalling that it is in agreement with the estimates obtained in 
the previous section, to within the errors. 

To evaluate the range of age differences that are still (at least marginally)
compatible with the data we explore the effects of the following (very unlikely)
occurrences: (a) we consider the maximum possible errors in matching the HB, as
evaluated in \S4.1, Figs.~4 and 5, all in the directions leading to the
maximum overestimate of the age differences, and (b) we consider the maximum
possible errors all in the opposite direction, leading to the maximum
underestimate of the age difference. The results are shown in Fig.~12 by
comparing the ridge lines with the SCL97 set of isochrones,
with the same arrangement and symbols as in Fig.~11. An additional shift 
has been also applied to the ridge lines, corresponding to the error bars of the
shifts adopted to match the HBs.
Case (a) is considered in panel {\em (a)} of Fig.~12. The additional shifts
$\delta(V-I)=+0.01$~mag and $\delta V=+0.01$~mag have been applied to the ridge line 
of NGC~362, such shifts being $\delta(V-I)=-0.01$~mag and $\delta V=-0.07$~mag  
in the case of NGC~288. The age difference is still $\sim 1$~Gyr, as judged 
from the luminosity of the horizontal part of the SGB. The case (b) is shown in 
panel {\em (b)}: the same shifts have been applied to the ridge lines, but with
opposite sign, and a maximum age difference of $\sim 3$~Gyr is clearly obtained.
Also in these cases if one rely only on $V_{TO}$ for his estimate, larger
age differences would be obtained.

\begin{figure*}[ht]
\figurenum{12}
\centerline{\psfig{figure=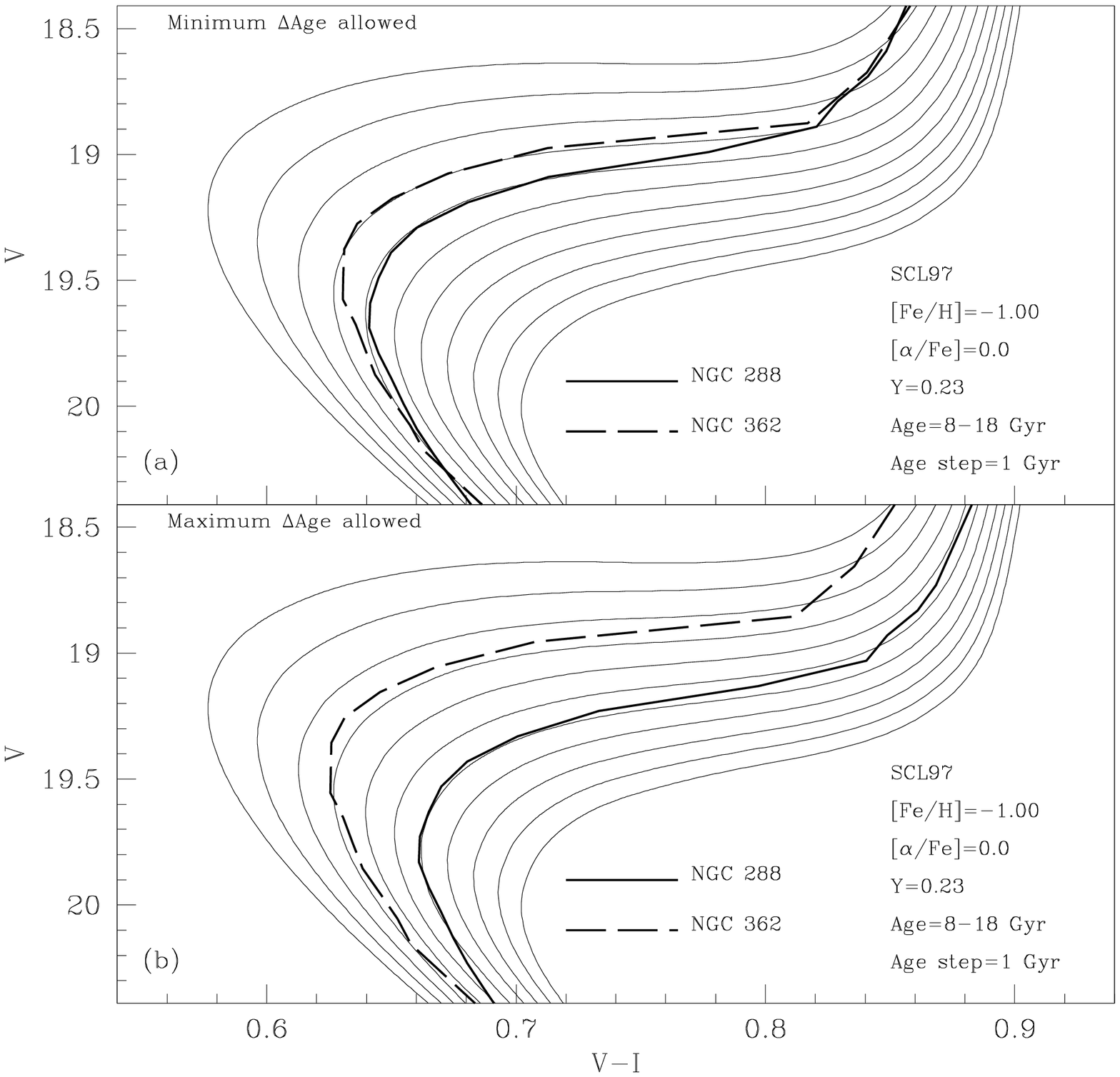}}
\caption{Comparison between ridge lines and isochrones, as in the 
panel {\em (a)} of Fig.~11. In the present case we have applyied the maximum
possible error in the shifts matching the HBs either all in one direction 
(panel {\em (a)}, bringing to the minimum age difference allowed by the data) 
or all in the opposite direction (panel {\em (b)}, bringing to the maximum age
difference allowed by the data).}
\end{figure*}

Formal errors of $\sim 0.5$~Gyr or lower are often associated with differential
age estimates. Despite the great reliability of the present result,
we prefer to provide a
conservative range of compatibility as the uncertainty associated with our age
estimate. Thus when we present the final result of our bridge test, i.e. {\em
the age difference between NGC~288 and NGC~362 is 
$\Delta {\rm Age}=2.0\pm 1.0$~Gyr}, what we are actually implying is 
that {\em age differences lower than +1~Gyr or larger than
+3~Gyr are excluded by the test}.

Some caveats associated to the assumptions of the bridge test will be discussed 
in \S4.6. 
 
\subsection{Direct comparison with previous bridge tests}

Following a suggestion of the Referee, we performed a comparison between
our photometry and the assembly of datasets adopted by SVB96 and V2000 in their
previous realizations of the bridge test. The comparison was required to clarify
the reasons for the different results we obtain with respect to SVB96
and V2000, given that the adopted shifts are roughly similar. 
The comparisons have
been made possible by the kind helpfulness of Dr. D.A.  VandenBerg who provided
the databases that he and SVB96 adopted in their tests.
Unfortunately the NGC~362 catalogue provided by Dr. VandenBerg was lacking
positional information, thus the cross-correlation with our data was not
possible.

In Fig.~13 we report: the difference between the V magnitudes measured in 
the present analysis and those by \citet{w92} - adopted by SVB96 and V2000 - 
versus our V magnitude for the common stars in NGC~1851 (panel {\em (a)}); 
for NGC~288 (panel {\em (b)}), the difference between our V magnitudes and 
those by \citet{berg93} (open triangles, adopted by SVB96 and V2000 for 
$V\le 17.5$), and by \citet{bol92} (full and open squares, adopted by SVB96 
and V2000 for $V< 17.5$) versus our V magnitude. The plots reported in panels 
{\em (c)} and {\em (d)} will be described below. The comparisons can be made
just for the V data since we have (V,I) photometry while SVB96 and V2000 based
their tests on (B,V) photometry.

The comparison of the photometry shown in Fig.~13 can be summarized as follows:

\begin{enumerate}

\item Panel {\em (a)}. Our V photometry and the one by \citet{w92} are in 
excellent agreement. Here the comparison is limited to $V\le 18.5$ because the
available catalogue lacks the faint stars. However this is the
only relevant range for NGC~1851 in the present test.

\item Panel {\em (b)}. The agreement with Bergbusch's photometry is quite good,
the average $\Delta V$ is $-0.016$ in the considered range of magnitudes.
However it is important to recall that both SVB96 and V2000 applied a shift of
$-0.06$ mag to report Bergbusch's data in the photometric system of
\citet{bol92}, following the prescriptions of the same
author\footnote{\citet{berg93} reports a shift of $+0.06$. We guess this is due
to a typographical error since both us and V2000 independently found that the 
true shift is in fact $-0.06$.}. 
Hence, the actual difference between our V photometry and the
photometry of bright stars adopted by SVB96 and V2000 amounts to $-0.076$ mag,
i.e. a quite sizeable mismatch.

\item Panel {\em (b)}. The \citet{bol92} dataset appears not consistent with
Bergbusch's photometry and also not self-consistent. This is not exceedingly 
surprising,
since this dataset was obtained by assembling different photometries,
obtained at different epochs, under different conditions and with different
instruments and telescopes. In particular the filled squares are stars measured
in the inner part of the cluster, with a photometric set-up mainly devoted to
the sampling of bright stars, while the open squares are stars from the outer, 
less crowded fields, observed with the aim of sampling the faintest part of 
the MS.
In the present plot we obtained the observed segregation just by plotting as
filled squares the stars with $X < 1600$ px and plotting as open squares the 
stars with $X > 1900$ px, in our reference frame. 
Such spatial threshold has the purpose of
showing separately the two different samples that are merged together in the
final dataset, putting in evidence the obvious unconsistency. Note that the
"faint" sample (open squares) naturally provides the clean MS shown in the CMD
presented in panel {\em (d)} (compare with panel {\em (c)}), and a measure of
$V_{TO}$ obtained from this sample would be brighter by $\sim 0.1$ mag, thus 
mimicking a younger age, more similar to NGC~362.

\item As an indirect test on the NGC~362 dataset we fitted our ridge lines
for NGC~362 and NGC~288 with the same isochrones and with the same adoption on
the apparent distance moduli as V2000. We obtained an absolute age of 
$\simeq 12$ Gyr for NGC~288 and $\simeq 11$ Gyr for NGC~362, i.e. broadly 
compatible with the results of V2000. This may be taken as an indication of
rough self-consistency between our dataset and the one adopted by V2000 for
NGC~362. Nevertheless, only  direct comparison would provide the final check.

\end{enumerate}

The above results provide a direct demonstration of the superiority of our 
bridge
test with respect to previous ones and clearly indicates the reasons of the
different result we obtain. We recall that our databases have been successfully
tested for self-consistency and linearity by the comparison with the photometry
by \citet{ros99}, performed {\em over the whole range of magnitude} 
covered by the data and in {\em both} passbands (see Fig.~2). 

\begin{figure*}[ht]
\figurenum{13}
\centerline{\psfig{figure=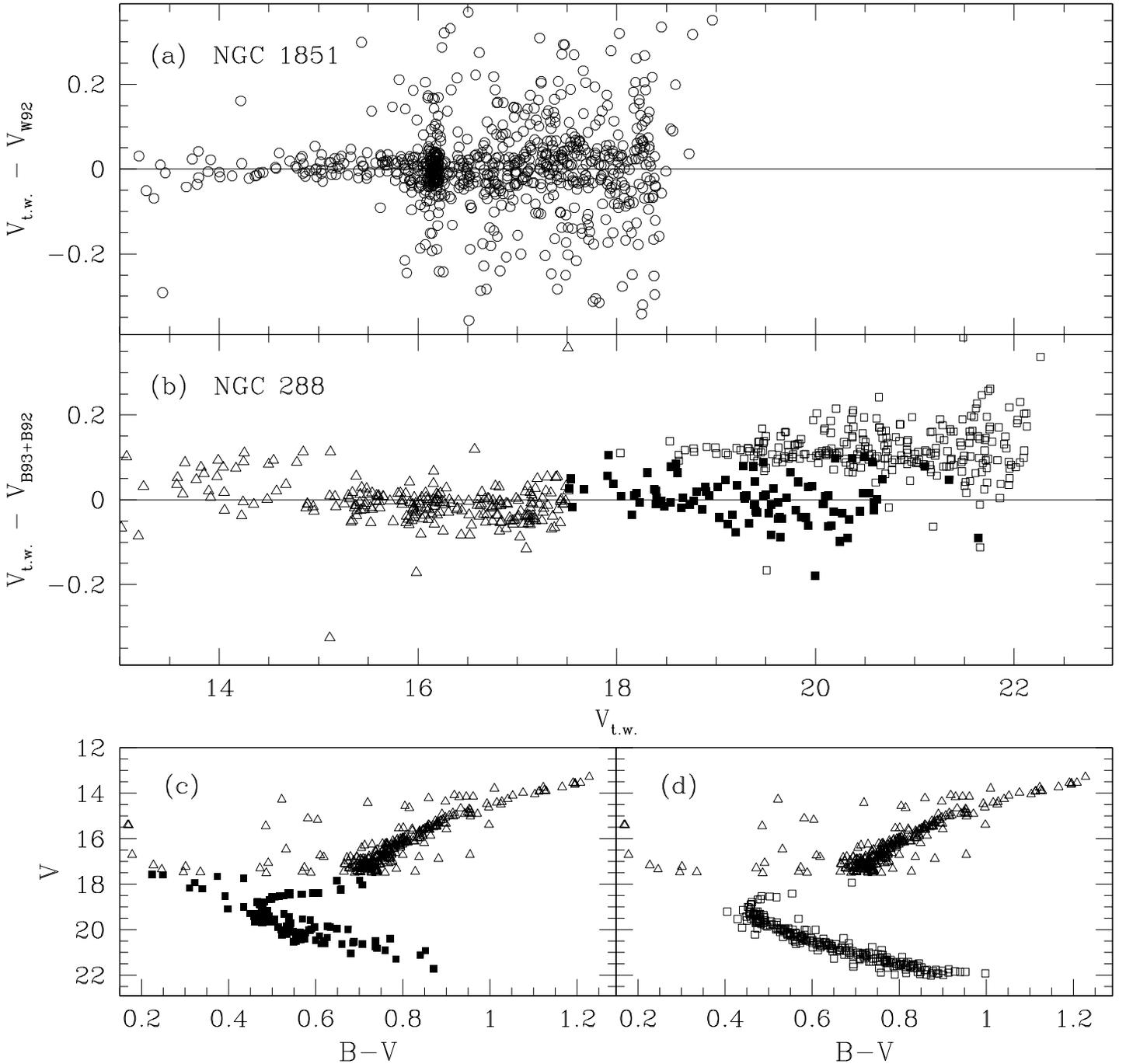}}
\caption{Comparison between our photometry and the datasets adopted by SVB96
and by V2000 in their previous realizations of the bridge test. 
Panel {\em (a)}: difference in V magnitude between our photometry and that by
\citet{w92} for NGC~1851, versus our V magnitudes ($V_{t.w.}$). 
Panel {\em (b)}: the same as panel {\em (a)} but for NGC~288. The open
triangles are the data from \citet{berg93}, the filled and open squares are
from the \citet{bol92} dataset. The filled squares are stars from the central
regions of the cluster (``bright'' sample, see text) while the open squares
are stars from the outer region of the cluster (``faint'' sample, see text).
The same symbols are adopted to show the different characteristics of the two
samples in the CMDs shown in panel {\em (c)} and {\em (d)}. Note that {\em all}
of the Blue Stragglers present in the dataset are in the ``bright'' central
sample, that have a limiting magnitude of $V\sim 21$ and a relatively small
number of MS stars . On the other hand the ``faint'' outer sample reaches
$V\sim 22$ and provides a very clean view of the TO region and of the MS. 
}
\end{figure*}

We want to stress here that the adoption of strictly homogeneous databases is
mandatory to obtain safe results from this kind of test 
\cite[see also][for further discussion on the importance of the homogeneity of
datasets]{ros99,ros00a}. This was the fundamental rationale at the basis of our
repetition of the bridge test (see \S2), and the above discussion provides a 
direct proof that the experiment was worth repeating. 

We emphasize that there are realistic cases where it is not possible 
to identify inconsistencies in 
a composite photometric datasets (e.g. few common stars, common stars in a 
restricted range of magnitude etc.). However one has to be aware of the danger
involved in the adoption of such datasets in such tricky business as the measure
of age differences.   

\subsection{Limitations of the Bridge Test}

The fundamental assumption at the basis of the bridge test is that {\em the
horizontal branch stars of clusters of the same metallicity ($Z$) and 
helium content ($Y$)
have the same luminosity at any given color}. In the present case, the first
underlying hypothesis, i.e. same metallicity, is clearly fulfilled while, as
already stated, 
we do not have fully conclusive constraints on $Y$ (see \S2.1 and \S4.5). A
higher helium content in NGC~288 would produce a brighter HB in this cluster,
thus mimicking an age difference. 
On the other hand, if NGC~362 were
He-enhanced, the age difference measured by the bridge test would underestimate
the actual age difference. In this regard we have to rely on the fact that the
observed $R$ parameters strongly suggest that the helium content of the considered
clusters is very similar. 

It may also be conceived that even if both these hypotheses
are fulfilled, the same {\em fundamental assumption} may be false, i.e. there is
some factor other than $Z$ and $Y$ that can differentially affect the HB 
{\em luminosities} of two clusters with similar $Y$ and $Z$ \citep{fure78}. 
The first possibility coming to mind is
{\em core rotation}, that may  make an HB star brighter and bluer than canonical 
expectations due, e.g., to an increase in the helium-core mass at the He-flash. 
It is interesting to note that the BHB stars of NGC~288 seem to be remarkably
fast rotators \citep{ruth}, while, unfortunately, no observational constraint 
in this sense is available for NGC~362 and NGC~1851. Recent measures of rotation
of HB stars in globulars \citep{behr} provided clear indications that we are far
from a complete understanding of the effects of rotation on HB stars. Hence, 
this possibility remains to be explored. We note that the $R$ method calibration 
{\em assumes} that the canonical He-core mass is the same for all clusters, 
which may not necessarily be true; independent observational constraints on 
the $M_{\rm c}$ value are difficult to obtain, as reviewed by \citet{cat96}. 

Deep mixing phenomena can alter the He 
content of some RGB stars, as proposed and discussed by \citet{swc}. The ``mixed
stars'' would place themselves on the ZAHB at higher $T_{\rm eff}$ and higher
luminosity with respect to ``non-mixed stars.'' If, for instance, this were
the origin of the bimodal HB of NGC~1851 \cite[i.e. non-mixed stars populating 
the RHB and mixed stars populating the BHB, as indeed suggested by][]{swc} 
then the adoption of its HB as the
bridge between NGC~288 and NGC~362 would lead to an {\em underestimate} of the 
{\em true} age difference. While there is no clear evidence of a difference in 
the deep mixing extent between NGC~288 and NGC~362 \citep{sk00}, 
we are unaware of sufficiently detailed spectroscopic analyses of NGC~1851 
bright giants that would conclusively rule out the possibility that its BHB 
stars are the progeny of He-mixed giants \cite[but see][]{bono}. 
At the same time, whether He mixing 
takes place at all is currently a much-debated issue, and we refer the reader 
to the papers by \citet{gru2,caval,we,gra01} for some recent insight on this 
complex problem. 

Finally, there is mounting evidence that many field blue subdwarf (sdB) stars
are in fact binary systems \cite[see][and references therein]{gls,sgb,maxt01},
supporting the possibility that some BHB stars may be the result of evolution
of some kind of binary \cite[see][and reference therein]{bai}. The consequences
of this scenario have not yet been studied and may affect our conclusion in some
unknown way. 

\begin{figure*}[ht]
\figurenum{14}
\centerline{\psfig{figure=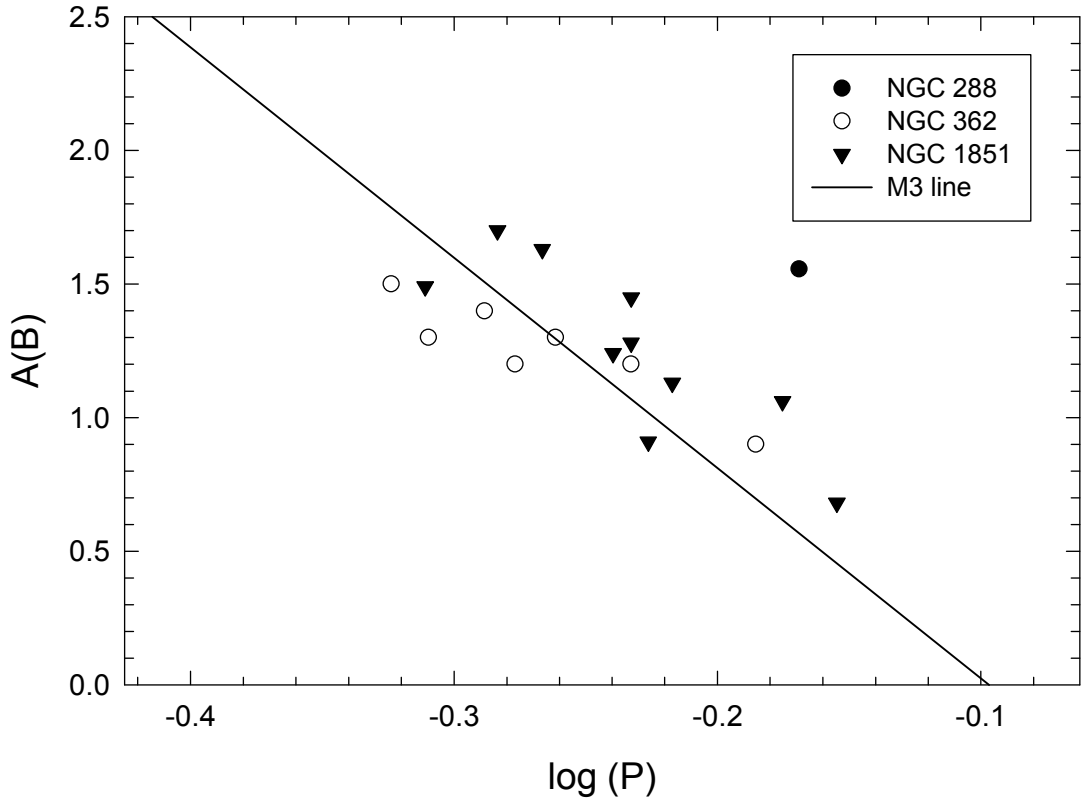}}
\caption{Bailey (period-blue amplitude) diagram for the fundamental-mode 
  RR Lyrae variables in NGC~1851 (inverted triangles), NGC~288 (filled circles) 
  and NGC~362 (open circles). The mean line for the M3 distribution, from 
  \citet{bori}, is provided as a reference. Note that there may be 
  a non-negligible period shift between NGC~1851 and NGC~362, and that the 
  NGC~288 RRab variable is clearly much brighter than {\em any} of the 
  variables in the other two clusters.}
\end{figure*}

One way to test the existence of luminosity differences between HBs 
of GCs having similar metallicity is through a traditional {\em period-shift 
analysis} of their fundamental-mode, RRab Lyrae variables. As pointed out by 
\citet{catel}, the pulsational properties of RR Lyrae 
variables can be useful indicators of a non-canonical origin for bimodal-HB 
and second-parameter clusters, given that these properties are sensitive to 
the basic physical parameters of the stars. In particular, differences in 
the HB luminosity should reflect themselves, at constant metallicity and 
temperature, in the form of noticeable ``period shifts" \cite[e.g.,][]{san90}.

Indeed, using this method, evidence has been reported that 
the NGC~1851 HB might be somewhat brighter than the HBs of 
other clusters of similar [Fe/H] \citep{catel}. 
However, \citet{w98} obtained new CCD data for the NGC~1851 RR Lyrae 
population and argued that the earlier evidence was a spurious consequence 
of the poor quality of the previously employed photographic data for this 
cluster's variables. 

The period-blue amplitude ($A_B$) diagram appears to be particularly suitable 
to carry out period shift analyses, in view of the evidence \citep{cat98,san00}
that $A_B$ is, at least to first order, 
a reasonable temperature indicator for the fundamental-mode RR Lyrae. 
In Fig.~14, we present this diagram for the RRab variables 
in NGC~1851, NGC~288, and NGC~362. The data for these three clusters were 
retrieved from \citet{w98,ka,clem}, respectively. Note that, in the 
case of Walker's data, we have discarded all stars which, according to his 
Fig.~3, may show signs of the Blazhko effect and/or have ill-defined light 
curves. For NGC~288, where only $V$ amplitudes are available, we obtained 
the $B$ amplitudes assuming 
$A_B/A_V \simeq 1.39$ \citep{layde}. In Fig.~14, we overplot, on the 
cluster data, the mean line obtained for the M3 (NGC~5272) RRab Lyrae 
variables by \citet{bori}. 

Figure~14 suggests that the NGC~1851 variables do have, at a given temperature 
(amplitude), systematically longer periods than their counterparts in NGC~362. 
Note, in particular, that while only 20\% of the NGC~1851 variables fall below 
the M3 line, as many as 57\% of the NGC~362 RRab Lyrae are located below such 
a line. The mean period shift between the two clusters, according to these 
data, amounts to $\Delta\log\,P (A_B) \approx 0.025$---which, if interpreted 
in terms of a difference in HB luminosity between NGC~362 and NGC~1851, implies 
that the latter has a brighter HB by $\approx 0.075$~mag. Though a variation in 
the RR Lyrae mass by $\approx 0.05\,M_{\odot}$ (with lower masses in the case 
of NGC~1851) could also account for such a period shift, it is unclear that 
any second parameter candidate could cause a change in mass {\em at a fixed 
effective temperature} without provoking a change in luminosity as well. 
Either option, or a combination of the two, presents problems for the use of 
NGC~1851 as the ``bridge" for the bridge test. Moreover, one will readily notice
from Fig.~14 that the single RRab in NGC~288 appears to be much brighter than 
any RRab stars in {\em either} NGC~1851 or NGC~362; therefore, in the event
that the NGC~1851 RR Lyrae overluminosity is interpreted as evidence of 
evolution away from a blue ZAHB, one must also conclude that, remarkably, 
{\em all} NGC~1851 variables are less evolved than the single RRab that is 
found in NGC~288. This, in turn, creates another potential problem for the 
bridge test, in the sense that the assumption that the blue HBs of NGC~288 
and NGC~1851 are entirely equivalent would break down. In fact, while some  
problems are encountered when attempting to model the redder BHB stars in 
NGC~288 (\citealt{vdb00}; Paper~II; Catelan et al., in preparation), the 
same effect, while also present, appears to be less severe in the case of 
NGC~1851. More data would be vital to solve 
these problems and place the bridge test on a firmer footing; in particular, 
spectroscopic gravities for the blue HB stars in both NGC~288 and NGC~1851 
and new RR Lyrae light curves for NGC~362 seem essential. 

Of course, if significant differences in helium content, 
core rotation and/or any other ``second parameter" that affects HB luminosity 
do exist and effectively change the HB luminosity of GCs with similar 
metallicity, our whole distance and age scales of GCs may be in error, 
since both are mostly based on the use of HB stars as standard candles 
\cite[e.g.,][]{pritz} -- the same fundamental assumption of the bridge test. 
Thus, while our present knowledge and data admittedly do not
allow complete control of all the variables that may affect a differential age 
estimate, and while some problems still exist that require further analysis, 
the {\em bridge test} is, with the currently available data, the most robust 
approach to estimate the age difference between NGC~362 and NGC~288 {\em within  
the canonical framework}. 

\section{The Origin of NGC~288 and NGC~362}

If an age difference of at least 2 Gyr between NGC~288 and NGC~362 is
confirmed, this will open a scenario for the formation of these clusters that
deserves some comment.
It is generally accepted that an $\alpha$-enhanced abundance pattern is the
signature of enrichment dominated by Type~II supernovae and, consequently, of a
short delay between the onset of star formation and the formation of the
$\alpha$-enhanced stars \cite[$< 1$ Gyr;][]{mcw97}. It is very interesting to
note that though NGC~362 seems to have formed $\sim 2$ Gyr later than NGC~288, 
both clusters are equally $\alpha$-enhanced. 
This suggests that both NGC~362 and NGC~288 were 
born shortly after the (local) onset of star formation, in regions/subsystems 
that had different evolutionary histories. The existence of subunits with
independent star formation and chemical enrichment histories in the early Galaxy
is consistent with the scenario envisaged by \citet{sz78} as well as with
modern Cold Dark Matter cosmological models.

\section{Summary and Conclusions}

We have performed an optimally
suited specific test to estimate the age difference
between the GCs NGC~288 and NGC~362. The bimodal HB of NGC~1851 
is used as a bridge to match the different HBs of NGC~288 and NGC~362 to a 
common level, so (hopefully) eliminating the effects of distance and reddening 
and providing the possibility of a direct comparison of the most age sensitive
features of the CMDs, i.e. the MSTO and SGB. We believe we have obtained a much
more robust estimate of the age difference between NGC~288 and NGC~362 with
respect to previous ones, for the following reasons:

\begin{enumerate}

\item We have adopted extremely homogeneous datasets, specifically tailored for
      the test;

\item A detailed and extensive abundance pattern comparison between NGC~288 and 
      NGC~362 was available for the first time \citep{sk00}, showing that
      the similarity in chemical composition between the two clusters is not
      limited to the overall metallicity but includes also $\alpha$-elements 
      (as well as other chemical species);
      
\item The shifts adopted to match the HBs provide an excellent
      overall match also for the RGB and for the lower MS, while 
      any other set of shifts we tried produced significant mismatches in 
      these sequences (see Fig.~8; Fig.~10 and \S4.2; see also V2000 and
      \citet{gru99}).       

\item We demonstrated that the previous realizations of the bridge test were
      plagued by non-self-consistency of the photometry, due to the adoption 
      of very heterogeneous datasets (see \S4.5).    
      
\end{enumerate}

According to the bridge test we find that {\em NGC~362 is younger than 
NGC~288 by} $2.0 \pm 1.0$~Gyr, in good agreement with the estimates we 
have also obtained from other differential age diagnostics 
$\Delta V_{\rm TO}^{\rm HB}$ and 
$\delta (V-I)_{@2.5}$, the latter being completely independent of the bridge 
test procedure. 

Our result is also in good agreement with the age scale recently obtained by 
\citet{ros99} from a very homogeneous photometric database, and is compatible 
with the previous findings by \citet{bol89,GN90,sd90}.
The possible sources of systematic error that may still affect our results are
associated with {\em (a)} the significant uncertainties in the $R$ parameters,
which leaves room for undetected differences in He content and {\em (b)} any
unknown (or unidentified) process able to significantly (and differentially) 
change the luminosity of the HB of at least one of the considered clusters. 
We critically discuss some possible caveats associated with the 
method utilizing, in particular, a period-shift analysis for the RR Lyrae 
variables in the clusters; this indicates some puzzling discrepancies and 
the need of better, modern time-series data for NGC~362 (see \S4.6).  

In a companion paper (Paper~II), the difference in HB morphology between 
NGC~288 and NGC~362 will be addressed in detail, in order to investigate 
anew whether our preferred age difference between NGC~288 and NGC~362 may 
completely account for the SPE in the considered case.

\acknowledgments 

This research has been funded by the Italian MURST through the COFIN p.
{\small 9902198923\_004} grant, assigned to the project 
{\em Stellar Dynamics and Stellar Evolution in Globular Clusters.}.
The financial support to S. Galleti has been provided by the Osservatorio
Astronomico di Bologna. Part of the data analysis has been performed using 
software developed by P. Montegriffo at the Osservatorio Astronomico di Bologna.
Support for M.C. was provided by NASA through
Hubble Fellowship grant HF--01105.01--98A awarded by the Space
Telescope Science Institute, which is operated by the Association
of Universities for Research in Astronomy, Inc., for NASA under
contract NAS~5--26555. M.C. thanks the Osservatorio Astronomico di Bologna
for the kind hospitality during the stage of revision of the present manuscript.

We are grateful to an anonymous Referee for having checked and analyzed very
carefully all the details of the present paper.

D.A. VandenBerg is acknowledged for providing his isochrones in electronic form,
the datasets needed for the presented comparisons, and for the many useful
discussions and suggestions.

We are grateful to M. Limongi, M. Castellani and S. Degl'Innocenti for their
help in the retrieving of isochrones sets from the ORFEO and GIPSY databases
respectively. We are grateful to F. Grundahl for communications of results
before publications and for useful discussions. D. Dinescu is warmly thanked 
for her help in the search for non member stars in NGC~288. 
This research has made use of NASA's Astrophysics Data System Abstract Service.

Dedicated to Beatrice Gualandi.



\clearpage

\begin{deluxetable}{lccccccc} 
\tablecolumns{8} 
\tablewidth{0pc} 
\tablecaption{Photometry of NGC 288. Selected sample.} 
\tablehead{\colhead{ID} & \colhead{$V$} & \colhead{$\epsilon _V$} &
 \colhead{$I$} & \colhead{$\epsilon _I$} & \colhead{$X_{\rm px}$} & 
\colhead{$Y_{\rm px}$}& \colhead{Other ID} \\}  
\startdata 
     1 & 13.068 & 0.012 & 11.605 & 0.009 &   270.54  &  735.19 & \nodata \\
     2 & 13.580 & 0.007 & 12.354 & 0.011 &   711.61  &  742.51 & \nodata \\
     3 & 13.709 & 0.021 & 12.704 & 0.028 &  1152.19  &  289.67 & \nodata \\
     4 & 13.654 & 0.010 & 12.406 & 0.012 &  1030.82  &  795.28 & \nodata \\
     5 & 14.956 & 0.015 & 13.969 & 0.021 &   246.91  &  315.11 & \nodata \\
     6 & 14.352 & 0.013 & 13.708 & 0.021 &  1440.71  & 1668.03 & \nodata \\
     7 & 16.289 & 0.035 & 15.442 & 0.036 &  1637.03  & 1680.05 & \nodata \\
\enddata 	  	 
\tablecomments{Table 1 is presented in its entirety in the electronic edition 
of the Astronomical Journal. A portion is shown here for guidance regarding its 
form and content.
Other ID: V2 and V3 are the RR Lyrae variables found by 
\citet{kal96}; Guo 4110 is a non member star identified in the proper motions 
database by \citet{guo}; EHB1 is the extreme HB star identified by \citet{bm99}.
} 	  	 
\end{deluxetable}


\begin{deluxetable}{lccccccc} 
\tablecolumns{8} 
\tablewidth{0pc} 
\tablecaption{Photometry of NGC 1851. Selected sample.} 
\tablehead{\colhead{ID} & \colhead{$V$} & \colhead{$\epsilon _V$} &
\colhead{$I$} & \colhead{$\epsilon _I$} & \colhead{$X_{\rm px}$} & 
\colhead{$Y_{\rm px}$}& \colhead{Other ID} \\}  
\startdata 
     1 & 15.917 & 0.010 & 14.912 & 0.010 &  1017.61 &	328.47 & \nodata  \\
     2 & 16.072 & 0.010 & 15.464 & 0.010 &   664.70 &	354.91 & V11 \\
     3 & 16.186 & 0.010 & 15.396 & 0.010 &   800.21 &	378.90 & \nodata  \\
     4 & 17.465 & 0.020 & 16.494 & 0.020 &  1239.68 &	455.88 & \nodata  \\
     5 & 16.165 & 0.010 & 15.344 & 0.010 &   492.83 &	515.76 & \nodata \\
     6 & 16.098 & 0.010 & 15.101 & 0.010 &  1202.77 &	590.08 & \nodata \\
     7 & 15.124 & 0.010 & 14.016 & 0.010 &  1043.44 &	659.45 & \nodata  \\
\enddata 	  	 
\tablecomments{Table 2 is presented in its entirety in the electronic edition 
of the Astronomical Journal. A portion is shown here for guidance regarding its 
form and content.
Other ID: stars from V2 to V28 are the RR Lyrae variables in common with
\citet{w98}; UIT-31 and UIT-44 are UV sources identified by the UIT satellite.
} 	  	 
\end{deluxetable}


\begin{deluxetable}{lccccccc} 
\tablecolumns{8} 
\tablewidth{0pc} 
\tablecaption{Photometry of NGC 362. Selected sample.} 
\tablehead{\colhead{ID} & \colhead{$V$} & \colhead{$\epsilon _V$} &
\colhead{$I$} & \colhead{$\epsilon _I$} & \colhead{$X_{\rm px}$} & 
\colhead{$Y_{\rm px}$}& \colhead{Other ID} \\}  
\startdata 
     1 & 14.805 & 0.004 & 13.758 & 0.005 &  1032.35 &	355.13 & \nodata \\
     2 & 15.358 & 0.005 & 14.482 & 0.010 &   946.28 &	361.07 & \nodata \\
     3 & 15.269 & 0.004 & 14.423 & 0.004 &   698.51 &	377.37 & \nodata\\
     4 & 16.708 & 0.005 & 15.321 & 0.005 &   748.67 &	398.34 & \nodata \\
     5 & 15.389 & 0.004 & 14.420 & 0.005 &   876.02 &	413.71 & \nodata \\
     6 & 15.502 & 0.005 & 14.739 & 0.006 &   847.30 &	423.78 & \nodata \\
     7 & 16.515 & 0.008 & 15.668 & 0.006 &  1116.93 &	429.80 & \nodata \\
\enddata 	  	 
\tablecomments{Table 3 is presented in its entirety in the electronic edition 
of the Astronomical Journal. A portion is shown here for guidance regarding its 
form and content.
Other ID: stars from V2 to V13 are the RR Lyrae variables identified in the
\citet{sh73} catalogue (version updated by C. Clement). PNM = probable non 
member (only for stars near the HB) according to \citet{t92}. MJ 6558 and MJ
8241 = BHB stars that are confirmed members of the clusters according to
\citet{m00}.} 	  	 
\end{deluxetable}


\begin{deluxetable}{ccccccccc} 
\tablecolumns{9} 
\tablewidth{0pc} 
\tablecaption{Average errors of the relative photometry} 
\tablehead{\multicolumn{1}{c}{}& 
\multicolumn{2}{c}{NGC 288} &   \colhead{}   & 
\multicolumn{2}{c}{NGC 1851} &   \colhead{}   &\multicolumn{2}{c}{NGC 362} \\ 
\colhead{$V$} & \colhead{$\epsilon_V$} & \colhead{$\epsilon_{V-I}$}   
& \colhead{}    & \colhead{$\epsilon_V$} & \colhead{$\epsilon_{V-I}$}      
& \colhead{}   & \colhead{$\epsilon_V$}    & \colhead{$\epsilon_{V-I}$}\\}
\startdata 
12.5 & 0.006 & 0.008 && \nodata&\nodata&& 0.006 & 0.010 \\
13.5 & 0.008 & 0.012 && 0.009 & 0.013 && 0.008 & 0.012 \\
14.5 & 0.009 & 0.013 && 0.006 & 0.011 && 0.007 & 0.009 \\
15.5 & 0.007 & 0.011 && 0.006 & 0.010 && 0.004 & 0.007 \\
16.5 & 0.006 & 0.010 && 0.006 & 0.010 && 0.007 & 0.009 \\
17.5 & 0.006 & 0.009 && 0.006 & 0.010 && 0.007 & 0.011 \\
18.5 & 0.008 & 0.013 && 0.004 & 0.007 && 0.011 & 0.018 \\
19.5 & 0.013 & 0.022 && 0.005 & 0.010 && 0.018 & 0.033 \\
20.5 & 0.024 & 0.042 && 0.009 & 0.017 && 0.034 & 0.060 \\
21.5 & 0.046 & 0.077 && 0.017 & 0.032 && 0.103 & 0.153 \\
22.5 & \nodata & \nodata && 0.043 & 0.077 && \nodata &\nodata\\
\enddata 	  	 
\end{deluxetable}


\begin{deluxetable}{cccccccc} 
\tablecolumns{8} 
\tablewidth{0pc} 
\tablecaption{MS and RGB ridge lines} 
\tablehead{ 
\multicolumn{2}{c}{NGC 288} &   \colhead{}   & 
\multicolumn{2}{c}{NGC 1851} &   \colhead{}   &\multicolumn{2}{c}{NGC 362} \\ 
\colhead{$V$} & \colhead{$V-I$}   & \colhead{}    & \colhead{$V$}  
& \colhead{$V-I$}    & \colhead{}   & \colhead{$V$}    & \colhead{$V-I$}\\}
\startdata 
12.6  & 1.814 &  & 13.3 & 1.818 &  & 12.6 &  1.609\\
12.9  & 1.492 &  & 13.4 & 1.609 &  &	13 &  1.424\\
13.15 & 1.374 &  & 13.80 & 1.391 &  & 13.5 &  1.292\\
13.65 & 1.262 &  & 14.15 & 1.297 &  &	14 &  1.189\\
14.15 & 1.154 &  & 14.65 & 1.188 &  & 14.5 &  1.079\\
14.65 & 1.089 &  & 15.15 & 1.112 &  &	15 &  1.018\\
15.15 & 1.026 &  & 15.65 &  1.05 &  & 15.5 & 0.9607\\
15.65 &0.9663 &  & 16.15 &0.9993 &  &	16 & 0.9206\\
16.15 &0.9337 &  & 16.65 &  0.96 &  & 16.5 & 0.8961\\
16.65 &0.9045 &  & 17.15 &0.9286 &  &	17 & 0.8653\\
17.15 &0.8795 &  & 17.65 &0.8979 &  & 17.5 & 0.8486\\
17.65 &0.8591 &  & 18.15 &0.8677 &  &	18 & 0.8157\\
   18 &0.8434 &  & 18.65 &0.8493 &  & 18.2 & 0.7919\\
 18.1 &0.8361 &  & 18.82 &0.8358 &  & 18.3 & 0.6876\\
 18.2 &0.8239 &  &    19 & 0.738 &  & 18.4 & 0.6477\\
 18.3 &0.8155 &  &  19.1 &0.6733 &  & 18.5 & 0.6255\\
 18.4 &0.7723 &  &  19.2 &0.6527 &  & 18.6 & 0.6112\\
 18.5 &0.7081 &  &  19.3 & 0.641 &  & 18.7 &  0.606\\
 18.6 &0.6757 &  &  19.4 &0.6367 &  & 18.8 & 0.6057\\
 18.7 &0.6553 &  &  19.5 &0.6322 &  & 18.9 & 0.6055\\
 18.8 &0.6449 &  &  19.6 &0.6321 &  &	19 & 0.6106\\
 18.9 &0.6399 &  &  19.7 &0.6349 &  & 19.1 & 0.6145\\
   19 &0.6364 &  &  19.8 &0.6393 &  & 19.2 & 0.6185\\
 19.1 &0.6361 &  &  19.9 &0.6444 &  & 19.3 & 0.6255\\
 19.2 &0.6399 &  &    20 &0.6493 &  & 19.4 & 0.6324\\
 19.3 & 0.645 &  &  20.1 &0.6548 &  & 19.5 & 0.6379\\
 19.4 &0.6498 &  &  20.2 &0.6614 &  & 19.5 & 0.6379\\
 19.5 &0.6553 &  &  20.3 &0.6682 &  &	20 & 0.6926\\
 19.6 &0.6619 &  &  20.4 &0.6747 &  & 20.5 &  0.749\\
 19.7 &0.6691 &  &  20.5 &0.6825 &  &	21 & 0.8254\\
 19.8 &0.6769 &  &  20.5 &0.6825 &  &	   &	   \\
 19.8 &0.6769 &  &    21 & 0.745 &  &	   &	   \\
 20.3 &0.7288 &  &  21.5 &0.8135 &  &	   &	   \\
 20.8 &0.8026 &  &    22 &0.8998 &  &	   &	   \\
 21.3 &0.8988 &  &	 &	 &  &	   &	   \\
 21.8 & 1.001 &  &	 &	 &  &	   &	   \\
\enddata 	  	 
\end{deluxetable}


\begin{deluxetable}{cccccccc} 
\tablecolumns{8} 
\tablewidth{0pc} 
\tablecaption{HB ridge lines} 
\tablehead{ 
\multicolumn{2}{c}{NGC 288} &   \colhead{}   & 
\multicolumn{2}{c}{NGC 1851} &   \colhead{}   &\multicolumn{2}{c}{NGC 362} \\ 
\colhead{$V-I$} & \colhead{$V$}   & \colhead{}    & \colhead{$V-I$}  
& \colhead{$V$}    & \colhead{}   & \colhead{$V-I$}    & \colhead{$V$}\\}
\startdata 

 -0.090 & 17.01 & & -0.030 &  16.89 & &       &      \\
 -0.045 & 16.24 & &  0.020 &  16.65 & &       &      \\
  0.019 & 15.93 & &  0.070 &  16.39 & &       &      \\
  0.109 & 15.54 & &  0.120 &  16.17 & &       &      \\
  0.243 & 15.39 & &  0.170 &  16.13 & &       &      \\
  	&	& &  0.220 &  16.11 & &       &      \\
	&       & &  0.270 &  16.06 & &       &      \\
\nodata &\nodata & \nodata & \nodata & \nodata & \nodata & \nodata & \nodata \\
	&       & & 0.650  &  16.17 & & 0.650 & 15.51\\
	&	& & 0.750  &  16.17 & & 0.750 & 15.51\\
	&	& & 0.850  &  16.17 & & 0.850 & 15.51\\
\enddata
\tablecomments{The first block of entries refers to BHBs, the second one to 
RHBs.}
\end{deluxetable}

\end{document}